\documentclass[aps,twocolumn,groupedaddress,superscriptaddress,amsmath,amssymb,prx]{revtex4-2}

\usepackage{bbold}
\usepackage{graphicx}
\usepackage{dcolumn}
\usepackage{bm}
\usepackage{bbm}
\usepackage{dsfont}
\usepackage{xcolor}
\usepackage{hyperref}
\usepackage{soul}
\usepackage{braket}

\begin{document}

\title{Non-Perturbative Floquet Engineering of the Toric-Code Hamiltonian and its Ground State}

\author{Francesco Petiziol} \email{f.petiziol@tu-berlin.de}
\affiliation{Technische Universit{\"a}t Berlin, Institut f{\"u}r Theoretische Physik, Hardenbergstraße 36, Berlin 10623, Germany}
\author{Sandro Wimberger}
\affiliation{Dipartimento di Scienze Matematiche, Fisiche e Informatiche, Università di Parma, Parco Area delle Scienze 7/A, 43124 Parma, Italy}
\affiliation{INFN, Sezione di Milano Bicocca, Gruppo Collegato di Parma, Parco Area delle Scienze 7/A, 43124 Parma, Italy}
\author{Andr{\'e} Eckardt}
\affiliation{Technische Universit{\"a}t Berlin, Institut f{\"u}r Theoretische Physik, Hardenbergstraße 36, Berlin 10623, Germany}
\author{Florian Mintert}
\affiliation{Blackett Laboratory, Imperial College London, London SW7 2AZ, United Kingdom}
\affiliation{
Helmholtz-Zentrum Dresden-Rossendorf, Bautzner Landstraße 400, 01328 Dresden, Germany
}

\date{\today}

\begin{abstract}
We theoretically propose a quantum simulation scheme for the toric-code Hamiltonian, the paradigmatic model of a quantum spin liquid, based on time-periodic driving. We develop a hybrid continuous-digital strategy that exploits the commutativity of different terms in the target Hamiltonian. It allows one to realize the required four-body interactions in a nonperturbative way, attaining strong coupling and the suppression of undesired processes. In addition, we design an optimal protocol for preparing the topologically ordered ground states with high fidelity. A proof-of-principle implementation of a topological device and its use to simulate the topological phase transition are also discussed. The proposed scheme finds natural implementation in architectures of superconducting qubits with tuneable couplings. 
\end{abstract}

\maketitle

\section{Introduction}

The discovery of the fractional quantum Hall effect~\cite{Tsui1982}, together with later theoretical formulations of quantum spin liquids~\cite{Kalmeyer1987}, unraveled the existence of phases of matter, eluding traditional classifications based on symmetry breaking and local order parameters, which can be understood instead through the notion of topological order~\cite{Wen1989, Wen1990}. Topologically ordered states are characterized by properties such as long-range entanglement, ground state degeneracy on topologically nontrivial manifolds, intrinsic robustness to perturbations, and they support quasiparticles with anyonic quantum statistics~\cite{Wen2017, Wen2004}. Such features, alongside their importance from a fundamental condensed-matter physics perspective, make topologically ordered states promising candidates for scalable quantum information processing~\cite{Fowler2012, Nayak2008, Terhal2015}. 

The paradigmatic model in the understanding of topological order and of its potential for quantum computation is Kitaev's toric-code Hamiltonian~\cite{Kitaev2003}. This model describes spin-1/2 systems in a two-dimensional (2D) square lattice, experiencing purely four-spin interactions. It underpins surface codes for fault-tolerant quantum computing~\cite{Bravyi1998, Dennis2002, Fowler2012} and is tightly bound to $\mathbb{Z}_2$ lattice gauge theory~\cite{Wegner1971, Kogut1979, Hamma2008, Savary2016} and string-net condensation~\cite{Wen2003b, Wen2004}. 
The development of quantum simulators and quantum processors has enabled to probe signatures of $\mathbb{Z}_2$ topological order in experiments~\cite{Lu2009, Luo2018, Song2018, Andersen2020, Semeghini2021, Satzinger2021}. For example, the toric-code ground state has been recently prepared in a superconducting quantum processor~\cite{Satzinger2021} and in neutral atom arrays~\cite{Bluvstein2022} via a quantum-circuit-based approach~\cite{Liu2022}, without implementing the background Hamiltonian. 
However, due to the difficulty to attain clean four-spin interactions in synthetic systems~\cite{Dai2017, Paredes2008, Weimer2010, Zohar2017, Sameti2017}, the realization of the full Kitaev's model is still a challenge. Realizing the toric-code Hamiltonian, beyond its ground state, is appealing for several reasons, both from the perspective of quantum information processing and to study topological quantum matter. First, it can provide the simplest self-correction mechanism, since the ground space becomes energetically favoured and protected against perturbations by the presence of a gap. Second, it makes the concept of anyonic quasiparticles well defined~\cite{Jiang2008}, as long-lived and localized collective excitations of the system. In a quantum computation, where excitations are errors, implementing a toric-code Hamiltonian with disordered couplings can localize them~\cite{Wootton2011, Stark2011}, preventing them to disperse in response to perturbations and thus facilitating their removal. Third, the Hamiltonian gives access to time evolution, thus enabling the investigation, for instance, of quench dynamics and related phenomena, such as entanglement growth and dynamical phase transitions~\cite{Srivastav2019, Heyl2018}. Fourth, it allows for the quantum simulation of the topological phase transition~\cite{Trebst2007, Yu2008, Hamma2008b, Vidal2009, Tupitsyn2010, Dusuel2011}, and potentially of the confinement-deconfinement transition of the associated $\mathbb{Z}_2$ lattice gauge theory~\cite{Savary2016}.

\begin{figure}[b]
\includegraphics[width=\linewidth]{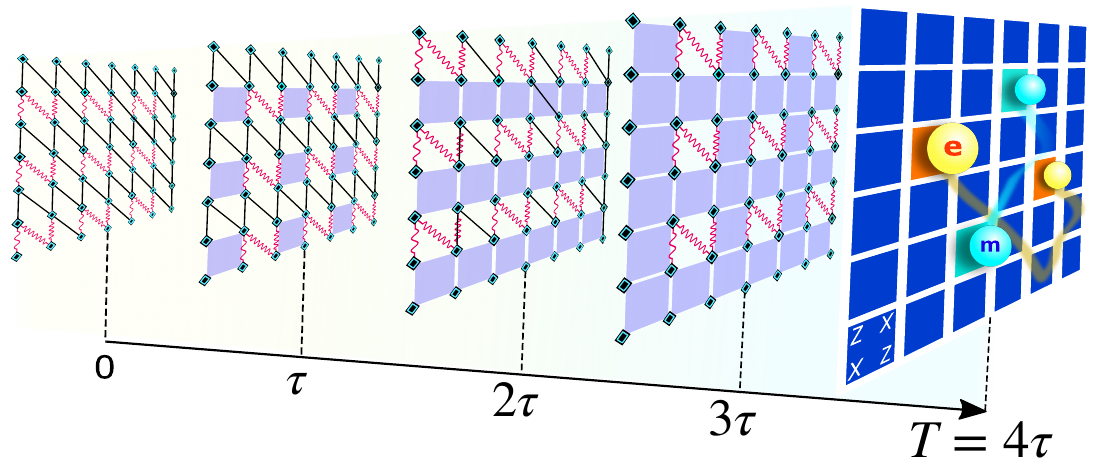}
\caption{Sketch of the spin lattice and of the quantum simulation approach, which will be detailed in the following. Square symbols indicate spins, solid lines represent tuneable pairwise couplings, and wavy purple lines represent couplings that are driven in a given time substep $\tau$, with associated angular frequency $\omega=2\pi/\tau$. In each substep $\tau$, different subsystems are decoupled from each other and driven to produce effective four-spin interactions (light-blue squares). At the end of the sequence (which is periodic with period $T=4\tau$), a complete toric-code Hamiltonian is achieved with high accuracy. The proposed scheme further allows one to prepare its ground state(s) and to study anyonic quasiparticles (`$\mathrm{e}$' and `$\mathrm{m}$').}
\label{fig:Figure1}
\end{figure}
In this work, we propose a scheme for the accurate quantum simulation of the toric-code Hamiltonian via periodic driving (sketched in Fig.~\ref{fig:Figure1}). Adopting the perspective of recently proposed hybrid approaches~\cite{Petiziol2021, Le2022}, we exploit the interplay of techniques from continuous-drive Floquet engineering~\cite{Goldman2014, Bukov2015, Eckardt2017, Oka2019} and trotterization~\cite{Georgescu2014} to minimize undesired terms in the quantum simulated Hamiltonian, achieving clean four-spin interactions. By taking advantage of the commutativity of different four-spin terms characterizing the Hamiltonian, we can reduce the problem of finding suitable control functions to individual four-spin subsystems, which can be optimized based on numerically exact methods, thus circumventing perturbative treatments. Moreover, this approach provides, by construction, a method that can be applied straightforwardly to arbitrary lattice size. Thanks to the fact that the proposed scheme exploits the integrability of the target model, the level of accuracy attained for the single plaquette is maintained when addressing the whole lattice: the leading error terms in the single plaquette remain the leading error terms also when considering the whole lattice, rather than being superseded by Trotter errors. It is further shown how the scheme can be adapted to implement four-spin entangling gates, which, in turn, allow us to design systematic ground-state-preparation protocols applicable in arbitrarily large systems. The achievement of topologically ordered states in these protocols is verified by probing long-range entanglement, detected via topological entanglement entropy, and through explicit creation and braiding of anyons. A minimal, proof-of-concept implementation is then proposed, which can be used either to study the transition to the ordered phase, through Floquet-adiabatic passage, or as a prototypical topological qubit.

Our scheme applies to a lattice of driven two-level systems and employs single-spin control and time-periodic modulations of the nearest-neighbour hopping. While such spin lattices are routinely realized in various quantum simulation platforms, the necessity to modulate spin-spin hopping makes architectures of superconducting qubits with tuneable-coupling a natural framework for implementing our proposal~\cite{Grajcar2006, Chen2014, Roushan2017, Yan2018, Arute2019, Weiss2022}. Indeed, this type of control has been demonstrated and used for the realization of efficient two-qubit gates \cite{Arute2019, Weiss2022} and artificial gauge fields for microwave photons \cite{Roushan2017}.

The presentation is organized as follows. In Section~\ref{sec:drive_ham}, the driven model and the target effective Hamiltonian are introduced. In Section~\ref{sec:single_plaquette}, the driving sequence which Floquet engineers the characteristic four-spin interactions of the toric code is presented. The protocol to generate the target Hamiltonian on the full lattice is then described in Sec.~\ref{sec:trotter}. In Section~\ref{sec:gs_prep}, it is shown how the toolbox developed can be used to prepare the toric-code ground state with high fidelity and, in Sec.~\ref{sec:anyons}, signatures of topological order in the state prepared are analysed. In Section~\ref{sec:prot_qubit}, we propose a minimal implementation of a proof-of-principle topological device and an adiabatic protocol to simulate the transition from a magnetically- to the topologically-ordered phase.

\section{Driven and target system Hamiltonians} \label{sec:drive_ham}

The Hamiltonian of the driven system describes two-level systems (spins) in a 2D lattice, and has the form $\hat{H}(t) = \hat{H}_1(t) + H_2(t)$ with
\begin{equation} \label{eq:Hdriven}
\hat{H}_1(t) = \sum_{\langle \alpha, \beta\rangle} g_{\alpha\beta}(t) (\hat{X}_\alpha\hat{X}_\beta + \hat{Y}_\alpha\hat{Y}_\beta)\ ,
\end{equation}
where the summation runs over nearest-neighbor pairs and $\{\hat{X}_\alpha, \hat{Y}_\alpha, \hat{Z}_\alpha\}$ denote Pauli matrices related to the $\alpha$th spin. The indices $\alpha$ and $\beta$ indicate pairs of coordinates $(i,j)$ in the 2D lattice. The geometry and connectivity of the lattice is sketched in Fig.~\ref{fig:Figure1}, and will be specified more in detail in the following (see Fig.~\ref{fig:Figure4}). The inter-qubit hopping will be periodically modulated in time, $g_{\alpha\beta}(t)=g_{\alpha\beta}(t+T)$. The Hamiltonian $\hat{H}_2(t)$ describes additional terms corresponding to resonant single-qubit pulses, of the form $\Omega_\alpha(t) \hat{\sigma}_\alpha$ for $\hat{\sigma}_\alpha=\hat{X}_\alpha,\hat{Y}_\alpha$ or $\hat{Z}_\alpha$. The Hamiltonian \eqref{eq:Hdriven} is written in the interaction picture with respect to the qubit energies (see Appendix \ref{sec:intpicH}) and can describe a lattice of superconducting qubits with controllable coupling, as has been realized with different strategies and architectures~\cite{Grajcar2006, Chen2014, Roushan2017, Yan2018, Arute2019, Weiss2022}. Modulation of $g_{\alpha\beta}(t)$ is achieved, for example, by means of an intermediate coupler and has been used to generated artificial magnetic fields~\cite{Roushan2017}. 

The target of the quantum simulation is to realize dynamics given by Wen's plaquette model \cite{Wen2003} on a square lattice,
\begin{align}  & \hat{H}_w = - \mathcal{J} \sum_{i,j} \hat{P}_{i,j} \ , \nonumber \\
& \hat{P}_{i,j}= \hat{X}_{i,j} \hat{Z}_{i,j+1} \hat{Z}_{i+1,j} \hat{X}_{i+1,j+1} \ .
\label{eq:wen}
\end{align}
Wen's model directly maps to the more commonly studied Kitaev's toric-code Hamiltonian~\cite{Kitaev2003} via a local basis rotation for a subset of qubits. Indeed, Kitaev's model is recovered by transforming Eq.~\eqref{eq:wen} through Hadamard gates $(\hat{X}_{i,j} + \hat{Z}_{i,j})/\sqrt{2}$ on all sites with $i+j$ either even or odd~\cite{Nussinov2009, Brown2011}. Besides, Wen's formulation has recently found renewed attention as a promising surface code design against biased noise~\cite{BonillaAtaides2021, Google2022}.
Our goal is then to find control functions $g_{\alpha\beta}(t)$ and $\Omega_\alpha$ such that the dynamics produced by the Hamiltonian~\eqref{eq:Hdriven}, at times matching multiples of a fundamental Floquet period $T$, reproduces as accurately as possible the dynamics generated by $\hat{H}_w$ of Eq.~\eqref{eq:wen}.

\section{Floquet engineering four-spin interactions} \label{sec:single_plaquette}

\begin{figure}[b]
\includegraphics[width=0.8\linewidth]{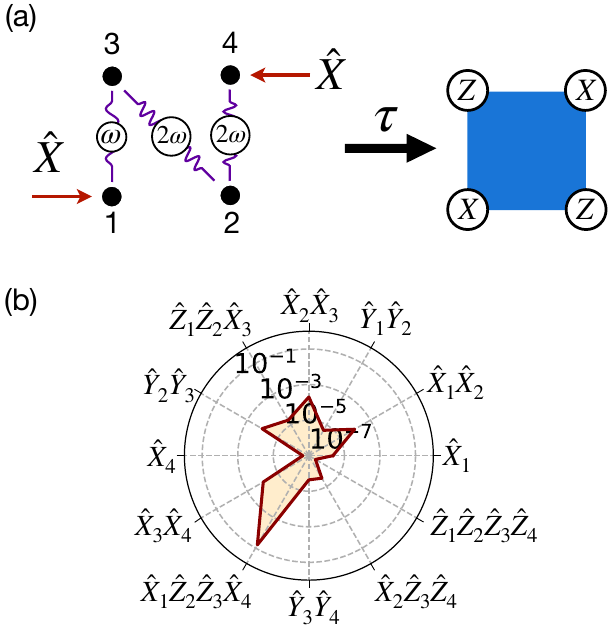}
\caption{(a) Sketch of the driving scheme for realizing a four-spin term in a single four-spin plaquette, involving oscillating fields at angular frequency $\omega$ and $2\omega$; (b) Components (larger than $10^{-7} \omega$ in magnitude) of the effective plaquette Hamiltonian, confirming the achievement of a clean four-spin interaction, in units of $\omega$.}
\label{fig:Figure2}
\end{figure}

The central difficulty in achieving the Hamiltonian $\hat{H}_w$ is the realization of the four-spin interactions $\hat{P}_{i,j}$ of Eq.~\eqref{eq:wen}, and the effective cancellation of natural single- and two-spin terms. This problem will now be addressed at the level of a fundamental four-spin plaquette. Considering a time step $\tau$ (which will be a submultiple of the overall Floquet period $T$), we analyze the effective Hamiltonian $\hat{H}_{i,j}$ generated at time $\tau$ by the driven Hamiltonian $\hat{H}(t)$ restricted to a single plaquette. The effective Hamiltonian is here defined by ($\hbar=1$)
\begin{equation}
e^{-i \hat{H}_{i,j} \tau} = \mathcal{T}\exp\left(-i\int_0^\tau \!\! dt \ \hat{H}(t) \right) \ .
\end{equation}
We then search for $\tau$-periodic control functions $g_{\alpha\beta}(t)$ yielding $\hat{H}_{i,j}\simeq- \mathcal{J}\hat{P}_{i,j}$ for some four-body coupling parameter $\mathcal{J}$.
 In Sec.~\ref{sec:trotter}, it will then be discussed how the resulting scheme can be used to engineer $\hat{H}_w$ on the full lattice, for an arbitrary system size. The starting point is a plaquette with coupling connectivity and spin labelling $k=1,2,3,4$ chosen as in Fig.~\ref{fig:Figure2}(a). The driven Hamiltonian of Eq. \eqref{eq:Hdriven}, in the absence of single-spin drives ($\hat{H}_2=0$), conserves the total magnetization $\sum_k \hat{Z}_k$ in the whole plaquette, while the target four-body term conserves only the total number of excitations residing on sites 2 and 3 instead.
This excitation conservation is broken for $\hat{H}(t)$ by introducing a strong resonant $\hat{X}$-drive on spins $1$ and $4$, $\hat{H}_2=  (\Omega_1 \hat{X}_1 + \Omega_4 \hat{X}_4)$ with amplitudes much larger than that of the two-spin coupling $g_{kk'}(t)$. The intuition is that this drive energetically penalizes processes induced by the $\hat{Y}\hat{Y}$, as compared to $\hat{X}\hat{X}$, interaction. These extra drive terms would not be needed if one is able to implement, separately, tuneable couplings of either $\hat{X}\hat{X}$ or $\hat{Y}\hat{Y}$ type. The Hamiltonian becomes
\begin{equation} \label{eq:drive_plaq}
\hat{H}(t) = \Omega_1 \hat{X}_1 + \Omega_4 \hat{X}_4 +\sum_{\langle k,k'\rangle} g_{kk'}(t) (\hat{X}_k\hat{X}_{k'} + \hat{Y}_k \hat{Y}_{k'})\ .
\end{equation}
In order to inspect the range of operators that can possibly compose the effective Hamiltonian generated by $\hat{H}(t)$, it is useful to study the dynamical Lie algebra $\mathfrak{L}$ of the control system~\cite{Dalessandro2007}. This is defined as the space spanned by all possible nested commutators of the control operators $\hat{X}_1$, $\hat{X}_4$, $\hat{X}_k\hat{X}_{k'} + \hat{Y}_k\hat{Y}_{k'}$ entering $\hat{H}(t)$. It characterizes entirely the set of evolution operators that can be produced by the driven dynamics and the set of reachable states~\cite{Dalessandro2007}. Hence, the effective Hamiltonian $\hat{H}_{i,j}$ for a plaquette can be generically expressed in the form
\begin{equation} \label{eq:Hdecomp}
\hat{H}_{i,j} = \sum_{\hat{O}_\ell\in \mathfrak{L}} c_{\ell} \hat{O}_\ell \ ,
\end{equation}
for some (unknown) coefficients $c_\ell$. A linearly independent set of operators $\hat{O}_\ell$ spanning $\mathfrak{L}$ is given in Table~\ref{tab:table1}. For each term, the number of commutators of the control operators needed to produce it is also reported.

\begin{table}[t]
\caption{\label{tab:table1}%
Basis of operators spanning the dynamical Lie Algebra of the single-plaquette system. The left column indicates the minimal number of nested commutators of initial control operators needed to produce the corresponding terms in the right column.
}
\begin{ruledtabular}
\begin{tabular}{l|c}
\# \textrm{commutators}&
\textrm{Operators}\\
\colrule
0 & $\hat{X}_1$, \, $\hat{X}_4$, \, $\hat{X}_1 \hat{X}_2+\hat{Y}_1\hat{Y}_2$, \\
& $\hat{X}_2 \hat{X}_3+\hat{Y}_2 \hat{Y}_3, \, \hat{X}_3 \hat{X}_4+ \hat{Y}_3 \hat{Y}_4$\\
\colrule
1 & $\hat{Z}_1 \hat{Y}_2$, \, $\hat{Y}_3 \hat{Z}_4$,\, $\hat{X}_1\hat{Z}_2 \hat{Y}_3 - \hat{Y}_1 \hat{Z}_2 \hat{X}_3$, \,  \\
& $\hat{X}_2\hat{Z}_3\hat{Y}_4 - \hat{Y}_2 \hat{Z}_3 \hat{X}_4$ \\
\colrule
2 & $\hat{Y}_1\hat{Y}_2$, \, $\hat{Y}_3\hat{Y}_4$, \, $\hat{Z}_1 \hat{Z}_2\hat{X}_3$,\, $\hat{X}_2 \hat{Z}_3 \hat{Z}_4$,\\
 &$\hat{X}_1 \hat{Z}_2 \hat{Z}_3 \hat{X}_4$,\, $\hat{Y}_1 \hat{Z}_2 \hat{Z}_3 \hat{Y}_4$, \,  \\
\colrule
3 & $\hat{Z}_1 \hat{Z}_2 \hat{Z}_3 \hat{Y}_4$, \, $\hat{Y}_1 \hat{Z}_2 \hat{Z}_3 \hat{Z}_4$ \\
\colrule
4 & $\hat{Z}_1 \hat{Z}_2 \hat{Z}_3 \hat{Z}_4$
\end{tabular}
\end{ruledtabular}
\end{table}

Building on intuition derived from high-frequency expansions for Floquet systems and explained in Appendix \ref{sec:plaquette_driving}, the driving functions are chosen as follows [and sketched in Fig.~\ref{fig:Figure2}(a)],
\begin{align} \label{eq:drive}
& g_{13}(t) = g_{13} \cos(\omega t)\  , \nonumber \\
&g_{23}(t) = g_{23} \cos(2\omega t)\ ,\nonumber \\
& g_{24}(t) = g_{24} \cos(2\omega t)\ ,
\end{align}
with angular frequency $\omega=2\pi/\tau$. The choice of functions which are time symmetric within one evolution period $\tau$, $g_{kk'}(t)=g_{kk'}(\tau-t)$, together with the particular structure of the dynamical Lie algebra, allows one to exclude from the effective Hamiltonian some of the terms belonging to $\mathfrak{L}$, as detailed in Appendix~\ref{app:lie}. 
 The remaining operators are linear combinations of those corresponding to an even number of commutators in Table~\ref{tab:table1}. 

Although the high-frequency regime and related traditional approaches based on high-frequency expansions are helpful for finding a promising ansatz for the functions $g_{kk'}(t)$, targeting the desired four-spin interactions directly with such methods poses several challenges. The four-body processes would appear at third order in a Floquet-Magnus expansion and have strength bounded by $g^3/\omega^2$ (Appendix \ref{sec:oscpart}), where $g$ is the maximal driving amplitude, which must satisfy $g\ll \omega$ for the expansion to converge. Then, on the one hand, the frequency $\omega$ should be kept small for the four-body coupling to be sizeable, but, on the other, it should be large to suppress higher-order terms in the expansion (and, in any case, to maintain $g/\omega \ll 1$ to ensure convergence). Even when a compromise between four-body coupling strength and accuracy has been accepted, the expansion up to third order will contain multiple other terms, with strength depending non-linearly on the driving amplitudes, making it very hard to single out the desired process only. To circumvent these limitations, we take advantage of the fact that the single-plaquette dynamics can be accessed in a numerically exact way~\cite{Petiziol2021}, thanks to the small system size. In particular, the driving amplitudes $g_{kk'}$ are numerically optimized together with $\Omega_j$ to achieve a four-spin interaction $\hat{X}_1 \hat{Z}_2 \hat{Z}_3 \hat{X}_4$ with a desired strength $\mathcal{J}$. This approach is nonperturbative, in the sense that it does not rely on high-frequency expansions to realize an effective Floquet Hamiltonian. Even though the strength of effective four-spin interaction is fundamentally limited by the strength of the actual two-body interactions [Eq.~\eqref{eq:Hdriven}],
the present approach can attain both strong and clean four-spin interactions at moderate driving frequency, since it combines effective processes to all orders.

As will be shown in detail in Sec.~\ref{sec:trotter}, where we use the single-plaquette engineering as part of the protocol to realize the full Hamiltonian for large systems, the error for engineering the Hamiltonian of the many-plaquette system will still be determined by (and of the same order of magnitude as) the single-plaquette error. This is a consequence of the fact that the single-plaquette terms of the target Hamiltonian commute with each other. Thus, the nonperturbative nature of the single-plaquette treatment allows also for a scalable nonperturbative engineering of the full toric code Hamiltonian. Details on the numerical optimization procedure are given in Appendix~\ref{sec:numerical_optimization}, and different optimizations can be performed to have a set of different values of $\mathcal{J}$ available. Since the Hamiltonian does not depend on $t$ and $\omega$ separately, but only on their product, in the following, $\omega$ is chosen as the unit of energy. 

The magnitude of different terms in the effective Hamiltonian of a single plaquette resulting from this procedure is shown in Fig.~\ref{fig:Figure2}(b), for a set of optimal parameters yielding a four-body term of chosen strength $\mathcal{J}\tau=\pi/8$. Approximated values of these parameters are 
given in Table~\ref{tab:Table2}. With this choice of $\mathcal{J}$, evolution for two Floquet periods produces an entangling four-spin gate $e^{i 2 \mathcal{J} \hat{P}_{i,j} \tau}= (\mathbb{1}+i\hat{P}_{i,j})/\sqrt{2}$, which will be employed in Sec.~\ref{sec:gs_prep} for ground state preparation.
\begin{table}[b]
\caption{Example sets of optimal driving parameters in units of $\omega$ (rounded to two significant figures). Row $A$ corresponds to a four-spin interaction strength $\mathcal{J}\tau=\pi/8$, while row $B$ corresponds to $\mathcal{J}\tau=\pi/50$. In row $C$, the parameters correspond to the three-spin interaction protocol described in Sec.~\ref{sec:prot_qubit} with $|\mathcal{J}_{xzx}|\tau=\pi/50$.}
\label{tab:Table2}
\begin{ruledtabular}
\begin{tabular}{l|lllll} 
 \multicolumn{6}{c}{four-spin interaction} \\
\colrule
& $\Omega_1$ & $\Omega_4$ &  $g_{13}$  &  $g_{23}$  & $g_{24}$  \\
\colrule
$A$ & 12 & 10 &  1.3 &  2.6 & 0.35\\
\colrule
$B$ & 14 & 11 & 1.2 & 2.6 & 0.054\\
\colrule
 \multicolumn{6}{c}{three-spin interaction} \\
\colrule
& $\overline{\Omega}_1$ &   $\overline{\Omega}_3$ & $g_{12}$ & $g_{23}$  & $[z_1, z_2, z_3]$ \\
\colrule
$C$ & 8.5 & 9.5 & 0.31 & 0.26 & [$-0.010$, 0.83, 2.4]
\end{tabular}
\end{ruledtabular}
\end{table}

As can be appreciated from Fig.~\ref{fig:Figure2}(b), the four-body interaction is successfully achieved and it is by far the dominant term in the effective Hamiltonian, with other terms almost three orders of magnitude smaller. The effective Hamiltonian for a plaquette can thus be written in the form
\begin{equation} \label{eq:H_ij}
\hat{H}_{i,j} = -\mathcal{J} \hat{P}_{i,j} + \varepsilon \hat{V}_{i,j} \ ,
\end{equation}
with a quantum simulation error $\varepsilon/\mathcal{J} \simeq 1.5\cdot 10^{-3}$ and with $\hat{V}_{i,j}$ containing the extra terms depicted in Fig.~\ref{fig:Figure2}(b) with dimensionless prefactors $c_\ell/\varepsilon < 1$. From Table~\ref{tab:Table2} one can observe that the value of the four-spin coupling considered $\mathcal{J}\approx 0.1\omega$ is around one order of magnitude smaller than the amplitudes of the oscillating drives $g_{kk'}$ and two orders of magnitude smaller than the maximal amplitude $\Omega_j$ of the static single-spin components $\hat{X}_1$ and $\hat{X}_4$. In a superconducting-circuit implementation, the latter can be taken one order of magnitude smaller than the qubit nonlinearities to avoid excitations outside of the qubit subspace. Considering that realistic dissipation rates in such platforms can be four-to-five orders of magnitudes smaller than such nonlinearities~\cite{Blais2021}, the four-spin interaction overcomes by one-to-two orders of magnitudes the dissipation and decoherence rates, thus attaining the strong-coupling regime.
 Let us remark that the bottleneck in the four-body coupling is given by the necessity to implement the $\hat{X}_k$ fields, which need to be larger than other driving parameters to break the conservation of excitations associated with the native flip-flop interactions $\hat{X}_k\hat{X}_{k'} + \hat{Y}_k\hat{Y}_{k'}$. If it is possible to directly oscillate {\it either} $\hat{X}_k\hat{X}_{k'}$ \emph{or} $\hat{Y}_k \hat{Y}_{k'}$, such terms would not be needed, and an order-of-magnitude gain in the four-spin interactions could be potentially achieved by increasing the driving frequency $\omega$ and amplitudes $g_{kk'}$.
 
\begin{figure}[t]
\includegraphics[width=\linewidth]{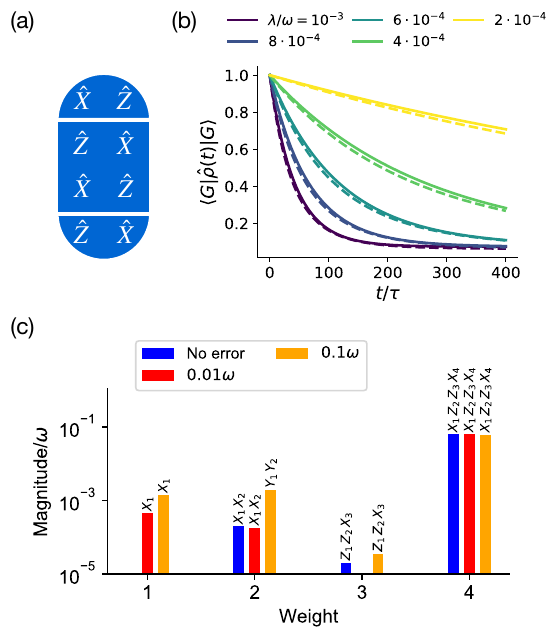}
\caption{(a) Two-by-two toric-code Hamiltonian with mixed boundary conditions. (b) Time evolution of the probability of remaining in the initial state $\ket{G}$, where $\ket{G}$ is a ground state of the toric code Hamiltonian shown in (a), for the undriven (dashed) and driven (solid) system in contact with thermal baths at temperature $k_BT=\omega_\alpha/15$, for different system-bath coupling values $\lambda$ and for $\omega_\alpha=2\cdot 10^3\omega$. (c) Sensitivity of the single plaquette protocol to errors in the control amplitudes. Different colors correspond to different maximal error amplitude $\eta_{\mathrm{max}}$. The bars represent the operator with maximal magnitude at a given weight (number of non-identity operators in the tensor product), which is indicated on the top of the bar.}
\label{fig:Figure3}
\end{figure}

Importantly, considering these parameter regimes, the presence of the Floquet drive is expected not to significantly alter decoherence and dissipation rates. Intuitively, this can be understood as follows. The Floquet protocol is defined in the interaction picture with respect to the bare spin transition energies $\omega_\alpha$ (Section~\ref{sec:drive_ham} and Appendix~\ref{sec:intpicH}) and the Floquet frequency $\omega$ is a small quantity as compared to $\omega_\alpha$. The environment couples to the spins inducing their relaxation (or, excitation), namely processes with an energy cost of $\omega_\alpha\gg\omega$. The Floquet drive is thus not able to provide or absorb sufficient energy to assist such processes and to significantly impact decay rates. To quantitatively verify this intuition, we compare the open-system dynamics of the single-plaquette system in the absence and in the presence of the drive. In Appendix ~\ref{app:master_equations}, Markovian quantum master equations are derived for the two scenarios, modelling the environment generically as a collection of bosonic thermal baths at realistic temperature, each coupling individually to one spin with coupling strength $\lambda$. For the driven system, the Floquet-Born-Markov formalism~\cite{Blumel1991, Grifoni1998, Mori2022} is employed, which captures potential transitions between Floquet states of the system assisted by the Floquet drive through the absorption or emission of integer multiples of driving quanta $m\omega$. The stroboscopic decay dynamics is depicted in Fig.~\ref{fig:Figure3}(b) for different values of the system-baths coupling strength $\lambda$, taken to reproduce realistic relaxation times $T_r$ for undriven superconducting qubits $T_r \omega_\alpha \sim 10^6$~\cite{Blais2021}. The figure of merit shown is the probability $\bra{G}\hat{\rho}(t)\ket{G}$ of remaining in the initial state $\ket{G}$ (Loschmidt echo) during the evolution, where $\ket{G}$ is taken as a toric ground state for mixed boundary conditions [depicted in~Fig.~\ref{fig:Figure3}(a)],
\begin{equation}
\ket{G}= \frac{1}{2}(1+\hat{Z}_1\hat{X}_2)(1+\hat{X}_3\hat{Z}_4) \ket{-11-},
\end{equation}
and where $\hat{\rho}(t)$ is the time-evolved density matrix of the system. The single-spin states $\ket{-}$ and $\ket{1}$ are eigenstates of $\hat{X}$ and $\hat{Z}$, respectively, with eigenvalue $-1$. The decay for the undriven system (dashed lines) is practically unaltered when introducing the Floquet protocol (solid lines), confirming that the latter does not induce any significant reduction of relaxation and decoherence times. The almost imperceptible increase of the decay time for the driven system can be attributed to the fact that the time-dependent drive weakly dresses the operators coupling to the baths, thus yielding a slight renormalization of the effective system-bath coupling---an effect exploited in the context of continuous dynamical decoupling~\cite{Fonseca2005,Viola1999, Lidar2014} as well as for reservoir engineering and nonlinearity control in driven transmon-cavity systems~\cite{Petiziol2022, Zhang2022}.

In order to benchmark the robustness of the effective four-spin Hamiltonian also against imperfections in the control parameters, the leading effective terms in the presence of errors are explored in Fig.~\ref{fig:Figure3}(c). In particular, errors are introduced in all parameters $p$ in the first row of Table~\ref{tab:Table2} according to $ p \longrightarrow p + \eta$, where $\eta$ represents uniformly distributed numbers in the interval $[-\eta_{\mathrm{max}}, \eta_{\mathrm{max}})$. The effective Hamiltonian is then averaged over 1000 realizations of errors affecting all parameters simultaneously. The resulting averaged effective Hamiltonian can be decomposed in the form \eqref{eq:Hdecomp} with strings of Pauli operators $\hat{O}_\ell$ and coefficients $c_\ell$.
Figure~\ref{fig:Figure3}(c) depicts the maximal value of the magnitudes $|c_\ell|$ of different terms for a given weight (number of non-identity terms in the Pauli string). One can appreciate that the effective four-body interaction persists even in the presence of relatively strong imperfections. For example, from Fig.~\ref{fig:Figure3}(c) one can see that errors larger than $0.1\omega$ must occur for the four-body term to be masked by other effective terms.

\section{Full lattice sequence} \label{sec:trotter}

\begin{figure}
\includegraphics[width=\linewidth]{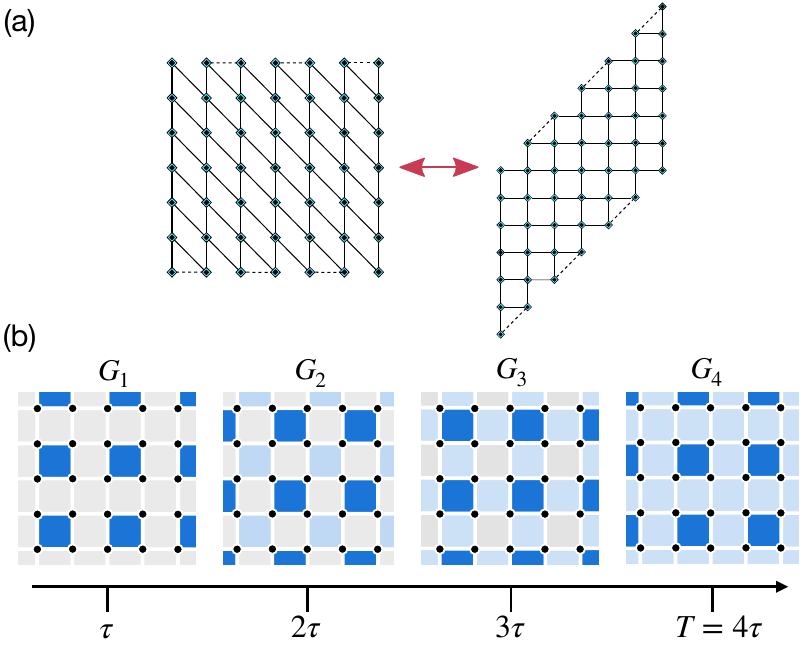}
\caption{(a) Lattice connectivity, equivalent to a distorted square lattice. Some triangular cells appear when including links (dashed lines) needed to implement mixed boundaries. (b) Trotter sequence to realize the Hamiltonian on the whole lattice. In each of the four steps of duration $\tau$, a group $G_k$ of disconnected plaquettes is driven (blue colour), while couplings connecting such plaquettes are turned off. The full lattice Hamiltonian progressively builds up (light blue).}
\label{fig:Figure4}
\end{figure}

The continuous-driving protocol for the single plaquette presented in Sec.~\ref{sec:single_plaquette} represents the fundamental building block that is used, in the following, to construct the full Hamiltonian on the whole lattice, by means of a Suzuki-Trotter sequence~\cite{Lloyd1996, Nielsen2004}, whose individual Trotter step will involve the continuous-drive protocol producing the four-spin interaction.  Although the present scheme can be applied to generic lattice topologies, for experimental practicality we will focus on planar lattices. The connectivity, scaling up the single-plaquette one, is shown in Fig.~\ref{fig:Figure4}(a), and is equivalent to a (distorted) square-lattice geometry. Some additional links [dashed in Fig.~\ref{fig:Figure4}(a)], making up triangular cells, are needed for implementing the mixed boundary terms discussed in the following.

A hybrid ``Floquet-Trotter'' quantum simulation strategy is particularly suited for the problem treated in this work and can attain a dramatic improvement as compared to straightforward digitization or high-frequency perturbative Floquet engineering. This is related to the interplay of two ingredients, namely, the fact that (i) at the single-plaquette level, continuous drives can be numerically optimized yielding very clean four-spin interactions within one time substep $\tau$, as discussed in Sec.~\ref{sec:single_plaquette}, and that (ii) such terms approximately commute on different sites, due to the properties of the toric-code Hamiltonian, thus yielding very small digitization errors when considering multiple plaquettes~\cite{Petiziol2021}. Hence, once the driving parameters for one plaquette are optimized, achieving the fundamental four-spin interaction with a small error $\varepsilon$ [see Eq.~\eqref{eq:H_ij}], scaling to arbitrary system size is immediate without dramatic error amplification due to digitization errors and without needing further parameter optimization. A direct trotterization would require instead a Trotter sequence to realize the four-spin interaction starting from native interactions, with related overhead of circuit depth, and would be plagued by the same limitations associated to high-frequency expansions discussed in Sec.~\ref{sec:single_plaquette}: the four-body coupling would be limited in coupling strength and accuracy by the perturbative estimate of the effective interactions based on Suzuki-Trotter formulae and the necessity to suppress higher-order terms via reduction of the Trotter step.

A possible first-order Floquet-Trotter sequence features four steps and is depicted in Fig.~\ref{fig:Figure4}(b). In each step, corresponding to one driving substep $\tau$, the couplings along lattice links connecting different blue-colored plaquettes are turned off by setting $g_{\alpha\beta}(t)=0$, while the single-plaquette protocol depicted in Fig.~\ref{fig:Figure2}(a) is applied to each dark-blue coloured plaquettes, engineering the desired four-body terms. These terms add up progressively in the sequence, building up the Hamiltonian of Eq.~\eqref{eq:wen} at time $T=4 \tau$. The propagator resulting from the Trotter sequence can be written in the form
\begin{equation} \label{eq:U_trotter}
\hat{U}(T) = \prod_{k=1}^4 \prod_{(i,j)\in {G}_k} \hat{U}_{ij}(\tau) \ ,
\end{equation}
where $\hat{U}_{ij}(\tau)=e^{-i \tau \hat{H}_{i,j}}$ represent single-plaquette propagators, while ${G}_k$ are the four groups of plaquettes involved in the sequence and depicted in Fig.~\ref{fig:Figure4}(b). Since each group $G_k$ contains disconnected plaquettes, the corresponding effective Hamiltonian is simply the sum of the effective Hamiltonians of the individual plaquettes involved,
\begin{equation}
\hat{H}_{k} = \sum_{(i,j)\in G_k} \hat{H}_{i,j} \ .
\end{equation}
The product of propagators $\hat{U}_k = e^{-i \hat{H}_k \tau}$ belonging to two different groups do not commute instead and can be estimated by means of the exponential product formula,
\begin{equation} \label{eq:expprodform}
\prod_{k} \hat{U}_{k} = \exp \left({-i \tau \sum_k \hat{H}_{k}+\frac{\tau^2}{2} \sum_{j<k} [\hat{H}_{j}, \hat{H}_{k}]+\dots}\right) .
\end{equation}
Using the fact that the desired terms $\hat{P}_{i,j}$ contained in $\hat{H}_k$ via the $\hat{H}_{i,j}$ commute with each other, the unwanted sub-leading-order terms in the exponential in Eq.~\eqref{eq:expprodform} are of the order of the error terms $\propto \epsilon$ in Eq.~\eqref{eq:H_ij}. The effective Hamiltonian at the end of the sequence can thus be estimated as
\begin{equation}
\hat{H}_{\mathrm{eff}} = \hat{H}_w + \widehat{\mathrm{Err}},
\end{equation}
where $\widehat{\mathrm{Err}}$ is at most of order $\varepsilon$,
\begin{align} \label{eq:Err}
& \widehat{\mathrm{Err}} = \varepsilon \sum_{i,j}\hat{V}_{i,j} -\\
&\frac{i \varepsilon\mathcal{J}\tau}{2} \sum_{k<k'}\sum_{\substack{(i,j)\in G_{k}\nonumber \\(i',j')\in G_{k'}}} \left([\hat{P}_{i,j}, \hat{V}_{i',j'}] + [\hat{V}_{i,j}, \hat{P}_{i',j'}] \right)+\dots \ . 
\end{align}
Thanks to the precision in the single-plaquette effective Hamiltonian achieved through numerical optimization, the Trotter error is very small. Thus, the integrability of the target Hamiltonian is used here in order to reduce errors in our hybrid Floquet-Trotter approach.

\begin{figure}[t]
\includegraphics[width=\linewidth]{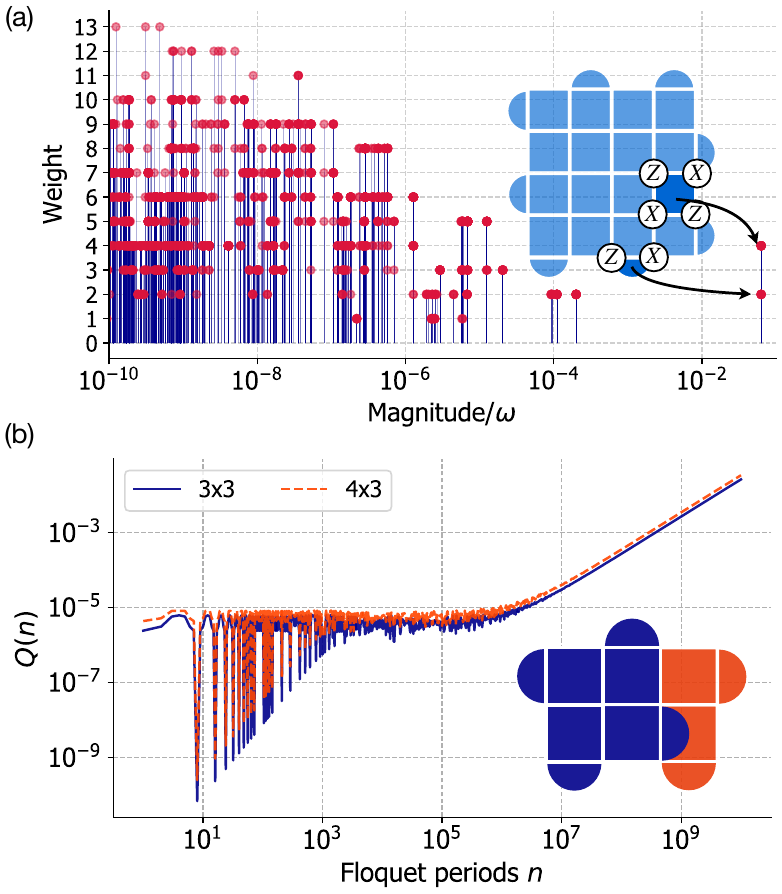}
\caption{(a) Weight vs magnitude $|c_\ell|$ of the operators composing the effective Hamiltonian for a 5-by-5 system. The desired four-spin and two-spin boundary terms (exemplified in the inset) are at least two orders of magnitude larger than undesired terms. The red colour of markers has non-unit opacity, such that more colour-intense markers appear as a result of overlapping markers. (b) Floquet heating dynamics, quantified via $Q(n)$ of Eq.~\eqref{eq:Qheat}, as a function of the number of Floquet periods $n$, for different system sizes with mixed boundary conditions as depicted in the inset.}
\label{fig:Figure5}
\end{figure}

To quantitatively inspect the magnitude of undesired terms arising from the combination of Floquet engineering and trotterization in a larger system, we consider a 5-by-5 lattice with mixed boundary conditions as depicted in the inset of Fig.~\ref{fig:Figure5}(a). Such a geometry is interesting, since it features two degenerate, topologically ordered ground states that can be used to define a topologically protected qubit~\cite{Satzinger2021}. The boundary terms, involving $\hat{X}_\alpha \hat{Z}_\beta$ terms, are Floquet engineered with the driving scheme
\begin{equation}
g_{\alpha\beta}\cos(\omega t)(\hat{X}_\alpha\hat{X}_\beta + \hat{Y}_\alpha\hat{Y}_\beta) + \tilde{g} \sin(\omega t) \hat{Y}_\beta\ ,
\end{equation}
where $g_{\alpha\beta}$ and $\tilde{g}$ are numerically optimized to give the same coupling $-\mathcal{J}$ as that engineered for the plaquette interactions in a time substep $\tau$.
The Floquet-Trotter realization of the Hamiltonian also allows us to develop a practical route to numerically compute the effective Hamiltonian for such a large system. Once the effective Hamiltonian $\hat{H}_{i,j}$ for a single plaquette (or boundary term) is determined numerically by diagonalizing the corresponding end-of-period propagator $\hat{U}_{ij}(\tau)=e^{-i \hat{H}_{i,j} \tau}$, we compute the full lattice Hamiltonian via Eq.~\eqref{eq:U_trotter} and iteratively using Baker-Campbell-Hausdorf formula for estimating the Hamiltonian, where terms in the series are included until convergence is reached. To avoid evaluating the commutators via explicit matrix multiplications, we develop a semi-analytic representation of the Hamiltonian (Appendix~\ref{sec:cpr_representation}), which allows one to compute the commutators analytically instead. 

Considering a decomposition of the Hamiltonian into Pauli strings with associated coefficients $c_\ell$, the magnitude $|c_{\ell}|$ of the different terms in the effective Hamiltonian is shown in Fig.~\ref{fig:Figure5}(a), where they are organized according to their weight (namely, the number of non-identity operators in the tensor product). The desired terms (four-body and boundary two-body) are almost three orders of magnitude larger than unwanted terms. The leading error terms appear at order $|c_\ell|/\omega\sim10^{-4}$ and correspond to two-spin $\hat{X}\hat{X}$ and $\hat{Y}\hat{Y}$ terms contained already in $\hat{V}_{i,j}$ [first line of Eq.~\eqref{eq:Err}]. The renowned topological properties of the toric code are in general jeopardized when strings of Pauli operators spanning the whole system and connecting different boundaries are present~\cite{Kitaev2003}. From Fig.~\ref{fig:Figure5}(a), one can further see that terms of such a weight are very weak, as expected since they can be created only via deeply nested commutators of single-plaquette effective Hamiltonians $\hat{H}_{i,j}$. For this reason, their magnitude will also decay with the system size, since more and more commutators will be needed to create a similar string. The preservation of the toric-code topological properties for the quantum simulated model will be inspected more in detail Sec.~\ref{sec:anyons}.

Before moving on, we would like to stress that our hybrid continuous-digital approach to Floquet engineering allows us to realize four-body terms, which in the regime where standard high-frequencies expansions work, would correspond to very small third-order contributions, as discussed in Sec.~\ref{sec:single_plaquette}. 

Next, we analyse the accuracy of the quantum simulation in terms of Floquet heating dynamics for the optimized protocol. Periodically-driven many-body systems are predicted to heat up to infinite temperature in the long-time limit, due to the absorption of energy from the drive~\cite{Dalessio2014, Lazarides2014,Moessner2017}, and it is thus interesting to characterize heating timescales induced by the error terms in the present model. With few exceptions~\cite{Lazarides2015,Abanin2016}, heating suppression requires large driving frequency, since heating processes are suppressed exponentially with increasing driving frequency~\cite{Abanin2015, Mori2016}, which is the premise of Floquet engineering in interacting many-body systems~\cite{Eckardt2015}. Here we show that, thanks to the nonperturbative approach to the four-spin interactions and the integrability of Wen's Hamiltonian, heating is strongly suppressed for the optimized driving parameters, without requiring very large driving frequency. To quantify heating, and thus also deviations from the ideal quantum simulation, we consider the long-time dynamics of the system initiated in the exact ground state $\ket{G}$ of Wen's Hamiltonian $\hat{H}_w$ [Eq.~\eqref{eq:wen}] and subjected to the Floquet-Trotter driving. We monitor stroboscopically the figure of merit~\cite{Heyl2019}
\begin{equation} \label{eq:Qheat}
Q(n) = \frac{\langle \hat{H}_w\rangle_n - E_0}{E_{\infty}-E_0},
\end{equation}
where $\langle \hat{H}_w \rangle_n = \bra{\psi(n T)} \hat{H}_w \ket{\psi(nT)}$ is the expectation value of Wen's Hamiltonian with respect to the stroboscopically time-evolved state $\ket{\psi(t)}$, $E_{\infty}=\mathrm{tr}[\hat{H}_w]/d$ is the energy of the infinite-temperature state with $d$ the size of the Hilbert space, $E_0 = \bra{G} \hat{H}_w\ket{G}$ is the ground-state energy. The quantity $Q(n)$ monitors deviations from the exact ideal dynamics, where $\hat{H}_w$ would be an exact constant of motion~\cite{Heyl2019}. Since $\hat{H}_w$ is traceless and its ground state energy is equal to the total number $N_s$ of stabilizer operators entering the Hamiltonian in units of $-\mathcal{J}$, $E_0 = -\mathcal{J} N_s$, Eq.~\eqref{eq:Qheat} reduces to $Q(n) = 1 + \langle \hat{H}_w\rangle_n/\mathcal{J} N_s $.
Long-time propagation is obtained via exact diagonalization of the one-period evolution operator $\hat{U}(T)$ and its exponentiation, which we can diagonalize up to a maximal size of 4-by-3 for the regular-rectangle geometries considered. The growth of $Q(n)$ as a function of the number of Floquet periods is depicted in Fig.~\ref{fig:Figure5}(b), for system sizes 3-by-3, which is the minimum size necessitating all the four Trotter steps of Fig.~\ref{fig:Figure4}(b), and 4-by-3, with mixed boundary conditions depicted in the inset of Fig.~\ref{fig:Figure5}(b). The heating measure $Q(n)$ remains constant and oscillatory for more than five decades--- i.e., well beyond timescales of practical applications--- after which the system starts to heat up. This heating timescale can be associated with the time it takes for error terms in the effective Hamiltonian to make themselves felt on the dynamics: since they have strength of the order of $10^{-4}\omega$ as visible in Fig.~\ref{fig:Figure5}(a) for the example of the 5-by-5 system, they require at least $\sim 10^5$ periods to have an appreciable impact. As predicted, the same behaviour is found for the both system sizes analyzed, confirming that the single-plaquette error terms remain the dominant imperfection also for larger systems.

\section{Ground state preparation} \label{sec:gs_prep}

By exploiting the possibility to realize the plaquette operators $\hat{P}_{i,j}$ with high precision and the related optimal driving parameters found, an efficient ground-state preparation protocol can be formulated, which can be applied to systems of arbitrary size. The realization of toric-code ground states, beyond the prominent interest in such states per se, is also emerging as a convenient preliminary step for addressing the preparation of more complex topologically ordered states~\cite{Kalinowski2022, Sun2022, Jin2022}. 

 The proposed protocol is optimal in the sense that the time needed to prepare the target state scales linearly with $N$ for an $N\times N$ lattice, thus saturating Lieb-Robinson bounds on the preparation of topological ordered states via local Hamiltonians~\cite{Bravyi2006}. It features a sequence in which different plaquette operators $\hat{P}_{i,j}$ are Floquet engineered in a way such that evolution for two time steps $\tau$ implements a four-spin entangling gate, with the help of additional single-qubit $\hat{Z}_{i,j}$ rotations. A ground state $\ket{G}$ of $\hat{H}_w$ satisfies the ``stabilizer'' constraint $\langle \hat{P}_{i,j}\rangle=1$ on all plaquettes operators, and can be written as an equal superposition of closed strings~\cite{Kitaev2003},
\begin{equation} \label{eq:gs_proj}
\ket{G} = \prod_{i+j \, \mathrm{ even}} \frac{(1+\hat{P}_{i,j})}{\sqrt{2}} \ket{\psi_0},
\end{equation}
where $\ket{\psi_0} = \mathcal{H}\ket{0}^{\otimes N}$, with $\mathcal{H}$ the simultaneous Hadamard gate $(\hat{X}_{i,j}+\hat{Z}_{i,j})/\sqrt{2}$ on odd sites $(i,j)$, namely such that $i+j$ is odd [the bottom-left spin has coordinate (1,1)]. The sum in Eq.~\eqref{eq:gs_proj} involves even plaquettes. Even (odd) plaquettes are defined as those whose bottom-left site is an even (odd) site. Note that the ordering of factors in the product does not matter since all factors commute. The ground state obtained has the same parity of non-contractible string operators along the $x$ or $y$ direction as $\ket{\psi_0}$. In this case, for example, $\bra{G}\hat{X}_L\ket{G} = \bra{\psi_0}\hat{X}_L\ket{\psi_0}=1$, where $\hat{X}_L$ is a string of alternating $\hat{X}_{i,j}$ and $\hat{Z}_{i,j}$ operators spanning the whole system along the $y$ direction, which commutes with the Hamiltonian $\hat{H}_w$.

The individual operators $(1+\hat{P}_{i,j})$ are not unitary, and thus cannot be achieved by a Hamiltonian evolution. Nonetheless, an equivalent result is obtained, starting from state $\ket{\psi_0}$, by employing a suitably ordered product of operators
\begin{align}
\hat{\mathcal{U}}_{i,j}^{A} = (1-i \hat{A}_{i,j})/\sqrt{2} = \exp(-i \pi \hat{A}_{i,j}/4)\ ,\nonumber \\
\hat{\mathcal{U}}_{i,j}^{B} = (1-i \hat{B}_{i,j})/\sqrt{2} = \exp(-i \pi \hat{B}_{i,j}/4) \ ,
\end{align}
where
\begin{align}
\hat{A}_{i,j} = \hat{X}_{i,j} \hat{Z}_{i,j+1} \hat{Z}_{i+1,j} \hat{Y}_{i+1,j+1}\ , \nonumber \\
\hat{B}_{i,j} = \hat{Y}_{i,j} \hat{Z}_{i,j+1} \hat{Z}_{i+1,j} \hat{X}_{i+1,j+1}\ .
\end{align}
The construction of this protocol is discussed in Appendix~\ref{app:gs_prot}. The ground state $\ket{G}$ is obtained, for example, by applying ordered products of $\hat{A}_{i,j}$ (or $\hat{B}_{i,j}$) operators along bottom-left to top-right diagonals of the lattice involving even plaquettes, which are ordered from bottom-left to top-right for $\hat{A}_{i,j}$, and vice versa for $\hat{B}_{i,j}$. The choice between using operators $\hat{A}_{i,j}$ or $\hat{B}_{i,j}$ for a given diagonal depends on the boundary: a boundary term must always contain one $\hat{Y}$ operator, and hence the right boundary (and thus the whole corresponding diagonal) necessitates operators $\hat{B}_{i,j}$, whereas the left boundary needs $\hat{A}_{i,j}$.

\begin{figure}[t]
\includegraphics[width=\linewidth]{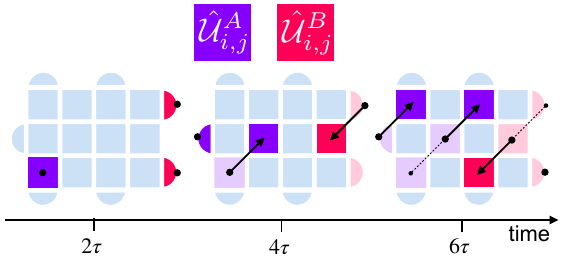}
\caption{Ground state preparation sequence for a 5-by-4 lattice. Dark and light purple tiles indicate an operation $\hat{\mathcal{U}}_{i,j}^{A}$ and $\hat{\mathcal{U}}_{i,j}^{B}$, respectively, each involving two time steps $\tau$ and two additional single-qubit rotations according to Eq.~\eqref{eq:rot_BC}. }
\label{fig:Figure6}
\end{figure}

Exploiting the Floquet-engineered plaquette operators, the gates $\hat{\mathcal{U}}_{i,j}^{A}$ and $\hat{\mathcal{U}}_{i,j}^{B}$ can be realized by combining the single-plaquette Floquet protocol with two single-qubit rotations. Indeed, introducing the notation $\hat{\mathcal{U}}_{i,j}^{P}=e^{i \frac{\pi}{4} \hat{P}_{i,j}}$ and $\hat{\mathcal{Z}}_{i,j}^{\pi/4}=e^{-i \frac{\pi}{4} \hat{Z}_{i,j}}$, we find
\begin{align} 
& \hat{\mathcal{U}}_{i,j}^{A} = \hat{\mathcal{Z}}_{i+1,j+1}^{\pi/4,\dagger} \hat{\mathcal{U}}_{i,j}^{P} \hat{\mathcal{Z}}_{i+1,j+1}^{\pi/4}\ , \nonumber \\
& \hat{\mathcal{U}}_{i,j}^{B} = \hat{\mathcal{Z}}_{i,j}^{\pi/4,\dagger} \hat{\mathcal{U}}_{i,j}^{P} \hat{\mathcal{Z}}_{i,j}^{\pi/4} \ . \label{eq:rot_BC}
\end{align}
To obtain the $\pi/4$ rotation needed, the driving parameters have been optimized to obtain $|\mathcal{J}| \tau= \pi/8$, such that evolution for two steps $2\tau$ implements the desired gate with $\pi/4$ angle. The single-qubit gates will typically be much faster than the time step $\tau$, such that one can approximately consider them to be instantaneous on the timescale of $\tau$. 

The sequence for a $5\times4$ lattice is shown in Fig.~\ref{fig:Figure6}. The state $\ket{G_{\mathrm{eff}}}$ approximating $\ket{G}$ prepared using the Floquet gates for such a system has an energy reaching 99.999\% of the exact toric-code ground-state energy and a fidelity $\left\lvert\braket{G_{\mathrm{eff}}|G}\right\lvert^2\simeq 99.994\%$. Since different diagonals can be addressed in parallel, the overall number of four-spin gates needed scales with the number of plaquettes tiling the longest diagonal of the lattice. This, in turn, scales linearly in the system size $N$, for a $N \times N$ system. Therefore, the ground state preparation protocol attains ``quantum speed limits'' on the preparation of topologically ordered states \cite{Bravyi2006}, being in this sense optimal. 

With respect to the protocol used in Ref.~\cite{Satzinger2021}, the protocol proposed here shares the strategy of preparing the ground state by realizing a unitary analogue of Eq.~\eqref{eq:gs_proj} starting from a trivial product state (at net of the mapping between Kitaev's and Wen's formulation of the model), though it uses a different gate sequence building on the four-spin gate. While both sequences scale optimally, the present method has the advantage that the realization of the Floquet gate can entangle four spins in a single evolution step, thus overcoming the necessity of multiple CNOT gates. For a comparison, a four-spin entangled state $(\ket{0+0+}+\ket{1-1-})/\sqrt{2}$ is created from $\ket{0+0+}$ in a single shot by the four-spin gate, but would require (at net of single-spin rotations) three CNOT gates instead using the method of Ref.~\cite{Satzinger2021}. Moreover, since the gate is based on an analogue Floquet-engineered four-spin coupling, it does not require further recompilation in terms of native interactions. These features may thus potentially yield a significant reduction of circuit depth for the ground state preparation.

\section{Probing entanglement and anyons} \label{sec:anyons}

The robustness of topological order in the toric code ground state against perturbations is the paradigmatic example of topological protection~\cite{Kitaev2003, Trebst2007,Hamma2008, Bravyi2010, Dusuel2011, Jamadagni2018}. Weak local perturbations can in general only slightly deform the ideal eigenstates, without spoiling their topological properties. Moreover, the ground state and anyonic excitations can still be efficiently manipulated by means of unperturbed string operators \cite{Bravyi2010}. The goal of this Section is to show that, despite the fact that the quantum simulation procedure also produces non-local higher-order contributions, approximate ground states prepared with the procedure described in Sec.~\ref{sec:gs_prep} indeed feature and enable the detection of topological order. To this end, typical signatures of topological order are probed in a 5-by-4 lattice, namely the presence of long-range entanglement and the existence of quasiparticles with non-trivial mutual and exchange statistics.

\begin{figure}[t]
\includegraphics[width=\linewidth]{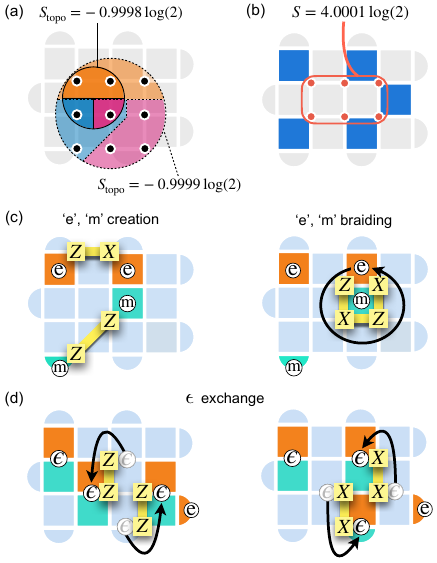}
\caption{(a) Topological entanglement entropy for two different partitions of the system, where the subsets $A$, $B$, $C$ are depicted in orange, purple and blue, respectively. (b) von Neumann entropy of the reduced state including the spins in red, which satisfies $(n-1)\log(2)$ where $n$ is the number of even plaquettes crossed (indicated in dark blue). (c) Specific anyon creation and braiding procedures and (d) dyon exchange, studied starting from the effective ground state $\ket{G_{\mathrm{eff}}}$ prepared through the Floquet scheme of Sec.~\ref{sec:gs_prep}.}
\label{fig:Figure7}
\end{figure}

To inspect the presence of entanglement as predicted in the toric code, the topological entanglement entropy $S_{\mathrm{topo}}$ \cite{Kitaev2006, Levin2006, Hamma2005} is reported in Fig.~\ref{fig:Figure7}(a) for two different partitions of the system. It is defined as~\cite{Kitaev2006}
\begin{equation}
S_{\mathrm{topo}} = S_A + S_B + S_C - S_{AB}-S_{BC}- S_{AC} + S_{ABC},
\end{equation}
where $S_X=\mathrm{tr} \rho_X \log \rho_X$ is the von Neumann entropy of the system state reduced to subsystem $X$. The subsystems $A$, $B$ and $C$ are depicted in orange, purple and blue color, respectively, in Fig.~\ref{fig:Figure7}(a). As indicated in the figure, $S_{\mathrm{topo}}$ is very close to the value of the ideal toric code, $-\log 2$, which characterizes $\mathbb{Z}_2$ topological order, and differs only by $\sim 10^{-4}$ in all cases. Additionally, it is verified that, for a single subsystem depicted in Fig.~\ref{fig:Figure7}(b), the von Neumann entropy scales like $(n-1)\log(2)$, where $n$ is the number of even plaquettes crossed~\cite{Satzinger2021}. 

As a second signature, the possibility to create and manipulate anyons in the same 5-by-4 system is probed, verifying mutual and exchange statistics through explicit braiding and exchange. The toric code possesses three types of gapped quasiparticle excitations, that appear as violations of the stabilizer constraint ($\langle \hat{P}_{i,j} \rangle=-1$) occurring at the end of open strings of Pauli operators. These are so-called electric charges ($`\mathrm{e}$') and magnetic vortices (`$\mathrm{m}$'), associated to a violation of even or odd plaquettes, respectively (the distinction is arbitrary~\cite{Bombin2010}), and dyons (`$\epsilon$', a compound $\epsilon = \mathrm{e}\times \mathrm{m}$ quasiparticle). Concerning the exchange statistics, `e' and `m' are bosons, whilst `$\epsilon$' are fermions. However, if an `$\mathrm{e}$' charge is braided around a `$\mathrm{m}$' charge or vice versa, the wavefunction acquires a $\pi$ phase, and thus these particles exhibit anyonic (semionic) mutual statistics. In the presence of weak perturbations, these quasiparticles are approximate eigenstates of the system, and will propagate with an effective finite mass~\cite{Kitaev2003}. Nonetheless, their characteristic properties are predicted to persist, until the perturbation becomes strong enough to disrupt the topological phase~\cite{Kitaev2003, Bravyi2010, Trebst2007, Dusuel2011, Savary2016}.

We verify these features through the processes depicted in Fig.~\ref{fig:Figure7}, where the creation and braiding of an electric charge around a magnetic vortex [\ref{fig:Figure7}(c)], and the exchange of two dyons $\epsilon$ [\ref{fig:Figure7}(d)], are depicted. These processes are realized in the system by applying strings of single-spin operators (shown in Fig.~\ref{fig:Figure7}) on top of the ground state $\ket{G_{\mathrm{eff}}}$ prepared with the protocol of Sec.~\ref{sec:gs_prep}. In the exact toric code, both operations result in the wavefunction acquiring a phase $\pi$ due to the semionic nature of `$\mathrm{e}$' and `$\mathrm{m}$', in the first case, and to the fermionic nature of `$\epsilon$', in the second. This is obtained almost exactly in the system studied here, where the phases obtained are $\pi$ up to numerical precision (resulting in a sign change $s$ deviating from $-1$ by $\sim10^{-4}$). The bosonic exchange statistics of `$\mathrm{e}$' and `$\mathrm{m}$' was also verified (not shown). 

These results give strong evidence that the state prepared is topologically ordered and features the defining properties of the $\mathbb{Z}_2$ topological phase very accurately. This further confirms that higher order contributions to the effective Hamiltonian have very little impact on the system properties, providing an additional signature of successful Floquet engineering.

\section{Proof-of-principle device and topological crossover}\label{sec:prot_qubit}

\begin{figure}[t]
\hspace{-0.5cm}\includegraphics[width=\linewidth]{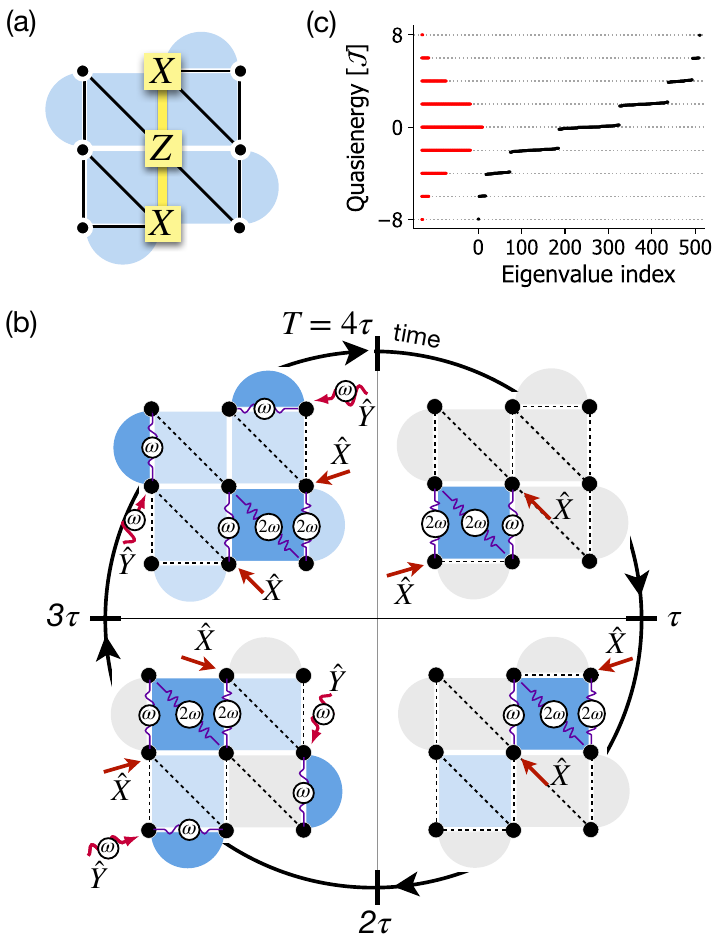}
\caption{(a) Device connectivity and $\hat{X}_L$ logical operator for the prototypical nine-spin device. (b) Floquet-Trotter sequence realizing $\hat{H}_{w}$. (c) Quasienergy spectrum of the effective Hamiltonian (black), compared to the exact spectrum of the ideal Hamiltonian (red).}
\label{fig:Figure8}
\end{figure}

In the earlier Sections, tools were developed for quantum simulating a toric-code Hamiltonian, preparing its ground state, and it has been verified that the prepared state indeed exhibits clear signatures of topological order. Based on this toolbox, we next propose a proof-of-principle realization, corresponding to a prototypical topological qubit, and show that it can be used to adiabatically explore the crossover into the spin-liquid phase. This (finite-size precursor of the) phase transition connects a topologically ordered to a topologically non-ordered phase and thus, unusually, it cannot be characterized through the behaviour of local order parameters. The proposed protocol provides a starting point for exploring the phase diagram of the perturbed toric code in a quantum simulator, the details of which are not yet entirely determined~\cite{Trebst2007, Yu2008,  Hamma2008b,Vidal2009, Tupitsyn2010, Dusuel2011}. This has potential impact also for the quantum simulation of lattice gauge theory, since the transition is related to confinement-deconfinement transitions of $\mathbb{Z}_2$ lattice gauge theory~\cite{Savary2016}. The prototypical device has the nine-spin geometry shown in Fig.~\ref{fig:Figure8}(a). The system features two degenerate ground states, which define a logical qubit. Topological order implies that these ground states cannot be distinguished via local observables, nor be turned one into the other by weak local perturbations. A string of single-spin errors needs to cross the entire system, involving at least three spins, to produce a logical error. The connectivity considered is shown in Fig.~\ref{fig:Figure8}(a), and the Floquet-Trotter sequence realizing the Hamiltonian, developed following the prescription of Sec.~\ref{sec:trotter}, is shown in Fig.~\ref{fig:Figure8}(b). The two ground states are distinguished by logical operators $\hat{X}_L$ and $\hat{Z}_L$ which span the system vertically and horizontally, commute with the Hamiltonian and anticommute with each other. A possible choice, which will be adopted in the following, is $\hat{X}_L = \hat{X}_{2,1}\hat{Z}_{2,2}\hat{X}_{2,3}$ and $\hat{Z}_L = \hat{Z}_{1,2}\hat{X}_{2,2}\hat{Z}_{3,2}$. Logical states $\ket{0_L}$ and $\ket{1_L}$ can then be defined according to
\begin{equation}
\hat{Z}_L\ket{0_L} = + \ket{0_L}, \quad \hat{Z}_L\ket{1_L} = - \ket{1_L} \ .
\end{equation}
For example, $\hat{X}_L$ is depicted in Fig.~\ref{fig:Figure8}(a). For a chosen four-spin coupling $\mathcal{J}=0.01\omega$, the quasienergy spectrum of the nine-qubit device is shown in Fig.~\ref{fig:Figure8}(c) [black colour], where it is also compared with the ideal spectrum (red colour). The quasienergy spectrum reproduces the ideal spectrum faithfully, featuring almost flat bands that acquire a slight dispersion due the weak higher-order terms in the effective Hamiltonian. 

While a system's ground state can be efficiently prepared with the protocol of Sec.~\ref{sec:gs_prep}, the goal here is to simulate the (finite-size precursor of the) topological transition by adiabatically interpolating between an effective Hamiltonian with topologically non-ordered ground state and Wen's model. Note that adiabatic---and even optimized diabatic~\cite{Raii2022}---preparation of topologically ordered states in interacting Floquet systems is an on-going challenge in state-of-the-art quantum simulators~\cite{Leonard2022}, due to the interplay of both nonadiabaticities and Floquet heating~\cite{Eckardt2005, Eckardt2017}, constraining applications to moderate system size.
We discuss how the Floquet-Trotter protocol proposed can be adapted to attain this goal and simulate the phase transition studied in Ref.~\cite{Yu2008}. The protocol aims at performing an interpolation from the initial effective Hamiltonian 
\begin{equation} \label{eq:Hinterp}
\hat{H}_{\mathrm{in}} = - R \sum_\alpha \hat{X}_\alpha + \hat{H}_w \ ,
\end{equation}
with $R \gg \mathcal{J}$, to a final effective Hamiltonian $\hat{H}_{\mathrm{fin}} = \hat{H}_w$. The topological phase transition for this model is predicted to occur at the critical point $R/\mathcal{J}\to 1$. For $R\gg \mathcal{J}$, the ground state is fully magnetized along the $x$ direction, and can be interpreted as a ``condensation" of both `e' and `m' quasiparticles~\cite{Yu2008}, while the spin-liquid ground state of $\hat{H}_w$ corresponds to condensation of closed strings. 
At the beginning of the protocol, the system is thus initialized with all spins aligned in the state $\ket{+}$, which is approximately the ground state of $\hat{H}_{\mathrm{in}}$. If the preparation is successful, at the final time the system will be in the topologically ordered ground-state-subspace of $\hat{H}_{\mathrm{fin}}$. 

The Floquet-Trotter sequence of Sec.~\ref{sec:trotter} implementing $\hat{H}_w$ is ``always-on'' during the protocol. The control parameters are optimized in order to achieve a weaker four-body coupling strength $\mathcal{J}=0.01\omega$: in this way, the quasienergy spectrum of $\hat{H}_w$ fits entirely within a single Floquet-Brillouin zone, simplifying the application of the adiabatic principle to the driven system~\cite{Eckardt2005, Eckardt2017}. The $\hat{X}_\alpha$ terms in $\hat{H}_{\mathrm{in}}$ are obtained by modifying the magnitude of single-spin terms in Eq.~\eqref{eq:drive_plaq} when driving each plaquette during the four-step sequence, $\Omega_k\to\Omega_k- R$. The interpolation takes place in a total time $t_f$ by ramping down $R\to R(t) = R r(t)$, with ramping function $r(t)$ such that $r(0)=1$ and $r(t_f)=1$. 

\begin{figure}[t!]
\includegraphics[width=\linewidth]{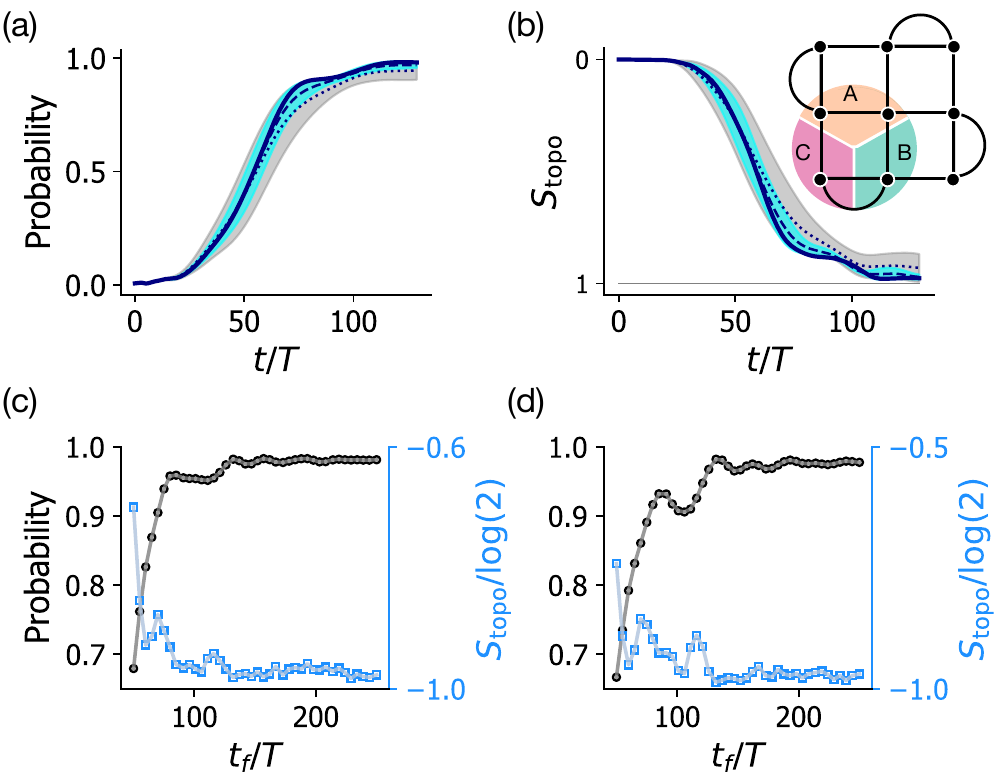}
\caption{Occupation probability of the ground-state-subspace (a) and growth of the topological entanglement entropy [(b), for the system partition shown in the inset] during the adiabatic ramp, for total evolution time $t_f=130T$. Dark-blue solid lines indicate the result using the optimized protocol, while the dotted and dashed lines indicate averaged results when including perturbations of such a protocol. These averages are obtained from 1000 realizations of random errors on the driving amplitudes, uniformly distributed in an interval of $\pm 1\%$ (dashed) and $\pm 2\%$ (dotted) around the optimized value. Light-blue and gray strips indicate a dispersion of one standard deviation from the dashed and dotted mean values, respectively. (c) Final ground-state-subspace occupation probability (black) and topological entanglement entropy (light blue) as a function of the total duration of the adiabatic interpolation. (d) Final occupation probability (black) and topological entanglement entropy (light blue) of the logical state $\ket{-_L}$ as a function of the total duration, for the adiabatic interpolation of Eq.~\eqref{eq:withXL}, which includes the logical operator $\hat{X}_L$. }
\label{fig:Figure9}
\end{figure}

An example of successful Floquet-adiabatic state preparation is obtained in $130$ Trotter steps using $R=10\mathcal{J}$, the sweep function $r(t)=\arctan(15 t/t_f)/\arctan(15)$ and is shown in Fig.~\ref{fig:Figure9}(a). In this Figure, the instantaneous squared overlap of the evolving state with the exact degenerate ground-state-subspace of the ideal toric code is reported (dark-blue thick solid line), which reaches a value $\simeq 0.98$ at the end of the sweep. The corresponding growth in time of the topological entanglement entropy, which reaches 98\% of the ideal value $-\log(2)$, is reported in Fig.~\ref{fig:Figure9}(b) (dark-blue thick solid line), for the system partition shown in the inset, and further signals the achievement of topological order. Final values of the ground space occupations and topological entanglement entropy at the end of the sweep for different protocol times are reported in Fig.~\ref{fig:Figure9}(c), while maintaining the same ramp. Occupations and $S_{\mathrm{topo}}$ above 95\% of the ideal values are attained for total durations $t_f \ge 80T$, after which the protocol is rather stable with respect to a variation in the total ramping time.

The adiabatic interpolation is robust also against imperfections in the Floquet drives. This is verified by perturbing the ideal Floquet-Trotter protocol during the ramp, by including random offsets in the driving amplitudes, $g_{ij}\to g_{ij} (1+\eta_{ij})$. The dashed and dotted lines in Fig.~\eqref{fig:Figure9}(a)-(b) represent the squared overlap averaged over $10^3$ realizations of uniformly distributed errors $\eta_{ij}\in[-\eta_{\mathrm{max}}, \eta_{\mathrm{max}})$, with $\eta_{\mathrm{max}}=1\%$ (dashed) and $2\%$ (dotted). The light-blue- and grey-coloured strips indicate a dispersion of one standard deviation around the mean value. The adiabatic interpolation is not spoiled by the presence of errors, and suffers only a moderate decrease in the final ground-state-subspace occupation, such that it still adequately realizes the topological crossover.

In the proposed nine-spin device, with an additional adaptation of the Floquet-Trotter sequence, it is also possible to precisely pinpoint a specific final state within the doubly degenerate ground-state-subspace, if desired. This can be done by including into the effective Hamiltonian a logical operator as follows,
\begin{equation} \label{eq:withXL}
\hat{H}_{\mathrm{in}} \to \hat{H}_{\mathrm{in}} \pm \hat{X}_L, \quad \hat{H}_{\mathrm{in}} \to \hat{H}_{\mathrm{in}} \pm \hat{X}_L\ .
\end{equation}
In this case, the final state will be a joint ground state of $\hat{H}_w$ and $\hat{X}_L$, which will be $\ket{+_L}$ or $\ket{-_L}$ depending on the choice of the sign in Eq.~\eqref{eq:withXL} and of the initial product state. Remarkably, the realization of the logical operator can be done via Floquet engineering in the prototypical nine-spin device, as discussed in the following. This becomes, however, prohibitively challenging for larger systems: since the logical operators need to span the whole system length, they will become higher and higher-order processes and thus become more and more challenging to be attained via two-body interactions and the Floquet drive.

Considering the three spins in the central column, labeled from 1 to 3 vertically from bottom to top, the protocol Hamiltonian involves periodic driving of the central site, adapting earlier protocols for achieving three-spin interactions~\cite{Petiziol2021}, and reads
\begin{align}
\hat{H}_{X_L}(t) =& \overline{\Omega}_1 \hat{X}_1 + \overline{\Omega}_3 \hat{X}_3 + g_{12}(\hat{X}_1\hat{X}_2 + \hat{Y}_1 \hat{Y}_2) \nonumber \\
&+ g_{23}(\hat{X}_2\hat{X}_3 + \hat{Y}_2 \hat{Y}_3) + z(t) \hat{Z}_2 \ .
\end{align}
The $\tau$-periodic drive $z(t)=z(t+\tau)$ is chosen to feature multiple harmonics,
\begin{equation} \label{eq:XLfloq}
z(t) = z_1 + z_2\cos(\omega t) + z_3 \cos(2 \omega t) \ .
\end{equation}
The parameters $\overline{\Omega}_j$, $g_{ij}$, $z_j$ are numerically optimized simultaneously to obtain an effective Hamiltonian $\mathcal{J}_{xzx}\hat{X}_L = \mathcal{J}_{xzx} \hat{X}_1 \hat{Z}_2 \hat{X}_3$, with a chosen magnitude $\mathcal{J}_{xzx}$. A set of optimal parameters is given in Table~\ref{tab:Table2}. The capability to implement a logical operator at the Hamiltonian level is also appealing, since it allows one to distinguish, and actively manipulate, protected qubit states. For example, it can be used to induce Rabi oscillations between the states $\ket{0_L}$ and $\ket{1_L}$. Here we focus on the adiabatic preparation of logical states $\ket{-_L}=(\ket{0_L}-\ket{1_L})/\sqrt{2}$, satisfying $\hat{X}_L\ket{-_L} = - \ket{-_L}$. The Floquet-adiabatic ramp used is the same, but each fundamental Trotter step now features one additional substep (thus, $T=5\tau$), which is used to implement $\hat{X}_L$ according to the driving pattern of Eq.~\eqref{eq:XLfloq}. The crossover into the target ground state is confirmed in Fig.~\ref{fig:Figure9}(d), where the final values of the squared overlap with state $\ket{-_L}$, and the topological entanglement entropy for the partition shown in Fig.~\ref{fig:Figure9}(b), are reported. The selected state is reached with probability larger than $95\%$ for protocol durations above $t_f=120T$.

For increasing system size $N$, the gap of the Hamiltonian of Eq.~\eqref{eq:Hinterp} at the critical point scales like $\sim1/N$~\cite{Yu2008}. This then indicates a linear scaling of the preparation time with the system size. However, the main difficulty in practically scaling up the adiabatic preparation can be expected to come from the interplay of the adiabatic ramp with the Floquet drives, producing Floquet heating. These effects become relevant when the quasienergy spectrum during the ramp spreads over multiple Floquet-Brillouin zones, such that accidental Floquet resonances (resonant $m$-`photon' processes borrowing $m$ energy quanta $\omega$ from the Floquet drive) are crossed producing unwanted excitations~\cite{Eckardt2017}. For the drive parameters used in Fig.~\ref{fig:Figure9}, this can be expected to happen for system sizes above 5-by-5, for which the spectral width of the Hamiltonian $\hat{H}_\mathrm{in}$ of Eq.~\eqref{eq:Hinterp} exceeds $\omega$. The error is expected to be marginal as long as the overlap between different Floquet-Brillouin zones is not substantial, since an overlap of the ground state with the most excited states is a high-order process and will lead to resonant excitation only on an exponentially long time scale. This can be further mitigated by optimization of the the sweep function~\cite{Eckardt2017}. These problems do not affect, instead, the state-preparation protocol presented in Sec.~\ref{sec:gs_prep} above, which relies entirely on the non-perturbative engineering of the Hamiltonian (and only prepares the final state without realizing the phase crossover studied here).

\section{Outlook}

The possibility to Floquet engineer a clean toric-code Hamiltonian as proposed in this work, with the added flexibility in the control of coupling and operators defining the effective Hamiltonian, provides a fully fledged quantum simulation strategy of $\mathbb{Z}_2$ lattice gauge theory~\cite{Savary2016, Hamma2008} and string-net condensation~\cite{Wen2003, Wen2003b,Wen2004}. Indeed, as exemplified in Section~\ref{sec:prot_qubit}, the Floquet-Trotter sequence can be easily adapted to incorporate additional single-spin terms, alongside the Floquet engineered four-spin interaction. By including local $\hat{X}_\alpha$ and $\hat{Z}_\alpha$ terms, the model can be interpreted as a $\mathbb{Z}_2$ lattice gauge theory with matter fields~\cite{Savary2016}. The tuneability of the different terms can then be exploited to explore, either via dynamical signatures or via adiabatic preparation, the phase diagram of such a model, and the crossover between the topological (or deconfined, or closed-strings-condensed~\cite{Yu2008}) and non-topological phase.

The proposed Floquet scheme provides also an interesting perspective for investigating quantum computational models related to the toric code. For example, it could be used to study quantum computation based on topological defects. Topological defects such as holes, which consist in the removal of a subset of stabilizer operators $\hat{P}_{i,j}$ from the Hamiltonian, admit the presence of anyons residing in the hole, which do not cost any energy (since the stabilizers in the hole have been removed from the Hamiltonian). 
The possibility to create, annihilate, and braid anyons, together with the holes potentially hosting them, enables universal quantum computation~\cite{Wootton2012}.
While anyon manipulation needs only single-spin operations, once the toric-code ground state has been prepared, creating and moving holes is typically a more challenging task~\cite{Wootton2012, Fowler2012}. This is particularly simple in the Floquet-Trotter scheme proposed here, where stabilizers $\hat{P}_{i,j}$ can be removed by simply avoiding to drive the corresponding plaquette in the sequence of Sec.~\ref{sec:trotter}. Similarly, a hole can be restored by simply reintroducing the corresponding drive in the subsequent Trotter step. To displace an occupied hole, one can combine such operations with single-spin operations to move the hosted anyon along with the hole. These fundamental operations then constitute a promising toolbox for implementing a prototypical Floquet quantum computation based on holes.

{\it Note added:} While finalizing this manuscript, other proposals for the Floquet engineering of quantum spin liquid models have appeared~\cite{Kalinowski2022, Sun2022}. Differently from the present paper, these works address the realization of Kitaev's honeycomb model for a non-abelian spin liquid~\cite{Kitaev2006b}, with focus on implementations with ultracold atoms~\cite{Sun2022} and Rydberg-atom arrays~\cite{Kalinowski2022}, and build on protocols based on van Vleck and Floquet-Magnus high-frequency expansions, respectively. 

\begin{acknowledgments} F.P. and A.E. acknowledge support from the Deutsche Forschungsgemeinschaft (DFG) via the Research Unit FOR 2414 under Project No. 277974659. S.W. acknowledges funding by the National Recovery and Resilience Plan (NRRP), Mission 4 Component 2 Investment 1.3 - Call for tender No. 341 of 15/03/2022 of Italian Ministry of University and Research funded by the European Union – NextGenerationEU, Project number PE0000023, Concession Decree No. 1564 of 11/10/2022 adopted by the Italian Ministry of University and Research, CUP D93C22000940001, Project title ``National Quantum Science and Technology Institute'' (NQSTI). \\
\end{acknowledgments}

\bibliography{biblio}

\cleardoublepage

\appendix

\section{Interaction picture Hamiltonian} \label{sec:intpicH}

The Hamiltonian of the system in the Schr\"odinger picture describes a lattice of two-level systems with tuneable nearest-neighbour hopping rate and subject to resonant qubit control,
\begin{align} 
\hat{H}(t) = & \sum_\alpha \Big[ \frac{\omega_\alpha}{2} \hat{Z}_\alpha + \Omega_\alpha(t) \sin(\omega_\alpha t+\phi_\alpha) \hat{Y}_\alpha \Big] \nonumber \\
& + \sum_{\langle \alpha, \beta \rangle}g_{\alpha\beta}(t) \cos(\omega_{\alpha\beta} t ) (\hat{\sigma}_\alpha^+\hat{\sigma}_\beta^- + \hat{\sigma}_\alpha^-\hat{\sigma}_\beta^+) \ , 
\label{eq:Hschro}
\end{align}
where $\hat{\sigma}_\alpha^\pm = (\hat{X}_\alpha \pm i \hat{Y}_\alpha )/2$, $\omega_{\alpha\beta} = \omega_\alpha - \omega_\beta$, $\Omega_\alpha(t)$ are the envelopes of the resonant qubit pulses, $g_{\alpha\beta}(t)$ a hopping modulation of frequency $\omega$, and $\langle \alpha, \beta\rangle$ indicates nearest neighbours. Indices $\alpha$ and $\beta$ denote pairs of lattice coordinates. The characteristic frequencies of $\Omega_\alpha(t)$ and $g_{\alpha\beta}(t)$ will be much smaller than the qubit frequencies $\omega_\alpha$ and their differences $\omega_{\alpha\beta}$. 
The Hamiltonian~\eqref{eq:Hschro} can describe an architecture of superconducting qubits with tuneable coupling and additional single-qubit microwave control, such as the one implemented in Ref.~\cite{Roushan2017}. In the interaction picture with respect to the perturbed problem, the Hamiltonian~\eqref{eq:Hschro} reads
\begin{align}
\hat{H}(t) = & \sum_\alpha  \frac{\Omega_\alpha(t)}{2} (e^{i\phi_\alpha} \hat{\sigma}_\alpha^+ + e^{-i\phi_\alpha}\hat{\sigma}_\alpha^- \nonumber \\
& -e^{i2\omega_\alpha t + i\phi_\alpha} \hat{\sigma}_\alpha^+- e^{-2i\omega_\alpha t -i\phi_\alpha}\hat{\sigma}_\alpha^- ) \nonumber \\
& + \sum_{\langle \alpha, \beta \rangle}g_{\alpha\beta}(t)  \big[\hat{\sigma}_\alpha^+\hat{\sigma}_\beta^- + \hat{\sigma}_\alpha^-\hat{\sigma}_\beta^+ \nonumber \\
& + e^{2i\omega_{\alpha\beta}t}\hat{\sigma}_\alpha^+\hat{\sigma}_\beta^- + e^{-2i\omega_{\alpha\beta}t}\hat{\sigma}_\alpha^-\hat{\sigma}_\beta^+\big] \ .
\end{align}
The second and fourth line involve terms oscillating at frequencies $\omega_\alpha$ and $\omega_{\alpha\beta}$, which are assumed to be much larger than the frequencies in $\Omega_\alpha(t)$ and $g_{\alpha\beta}(t)$. Hence, they can be neglected in rotating-wave approximation, and a Hamiltonian of the form of Eq.~\eqref{eq:Hdriven} is obtained. Note that, if single-qubit control was possible for resonant qubits, without crosstalk issues, the cosinusoidal modulation in Eq.~\eqref{eq:Hschro} could directly implement the modulations of Eq.~\eqref{eq:drive}, without needing the additional separation of timescales $\omega_{\alpha\beta}\gg\omega$.

\section{Single plaquette driving scheme}
\label{sec:plaquette_driving}

We motivate in this Section the choice of the driving functions of Eq.~\eqref{eq:drive}.

\subsection{Oscillating part}  \label{sec:oscpart}
 First, we explore the problem from the point of view of a high-frequency limit and related high-frequency expansions. This helps one to build some intuition about the choice of the driving fields, which provides the base for proceeding with numerically exact methods for designing the final protocol. The desired four-spin term $\hat{X}_1 \hat{Z}_2 \hat{Z}_3 \hat{X}_4$ can be written as the commutator of at least three two-spin operators,
\begin{equation} \label{eq:2comm}
\hat{X}_1 \hat{Z}_2 \hat{Z}_3 \hat{X}_4 \propto [\hat{X}_1 \hat{X}_2, [\hat{X}_2\hat{X}_3+\hat{Y}_2\hat{Y}_3, \hat{X}_3\hat{X}_4]]\ .
\end{equation}
Double-commutator terms of this form appear in a Magnus expansion ~\cite{Bukov2015, Eckardt2015, Eckardt2017} generated by the driven Hamiltonian $\hat{H}(t)$ at third order. In particular, such terms have, e.g., the form
\begin{align}
[ \hat{H}_{-m}, [\hat{H}_{m-n}, \hat{H}_n]]/\omega^2, \nonumber  \\
[\hat{H}_{-m},[\hat{H}_0, \hat{H}_m]]/\omega^2 \ , \dots \label{eq:2comm2}
\end{align}
where $\hat{H}_m = 1/\tau \int_0^\tau e^{i m \omega t} \hat{H}(t)$.
Choosing $\hat{H}_0=0$, since we want lower-order two-spin terms to be zero, leaves, for example, the first term. A choice of values of $m$ and $n$ that makes such a term non-zero is, for instance, $m=-1$, $n=2$. Then, comparing Eqs.~\eqref{eq:2comm} and \eqref{eq:2comm2}, one can start guessing a drive Hamiltonian of the form
\begin{equation} \label{eq:fguess}
g(\omega t) \hat{X}_1 \hat{X}_2 + g(-3\omega t) (\hat{X}_2 \hat{X}_3 + \hat{Y}_2 \hat{Y}_3) + g(2\omega t) \hat{X}_3 \hat{X}_4,
\end{equation}
with periodic $g(t+\tau)=g(t)$. Now, since the goal is to make third-order terms in the expansion to become of leading order, the second-order must vanish. This can be achieved by exploiting a property of the Magnus expansion, namely that all even order terms vanish if the driven Hamiltonian is time-symmetric within the integration step $\tau$, $\hat{H}(t) = \hat{H}(\tau-t)$~\cite{Burum1981, Brinkmann2016}. We will elaborate more on this point in subsection~\ref{app:lie}. It is thus convenient to choose $g(t)$ to be time-symmetric. We thus obtain driving functions (using the labelling of Fig.~\ref{fig:Figure2})
\begin{align} 
& g_{13}(t) = g_{13} \cos(\omega t) , \quad g_{23}(t) = g_{23} \cos(3\omega t) ,\nonumber \\
& \quad g_{24}(t) = g_{24} \cos(2\omega t),
\label{eq:ansatz_app}
\end{align}
which yield the desired four-body term with strength $\mathcal{J} \propto g_{13} g_{23} g_{24}/\omega^2$. The ansatz \eqref{eq:ansatz_app} is the starting point for the numerical optimization, where testing slightly different harmonics yields in the end optimal parameters with frequencies $\omega$ and $2\omega$ as in Eq.~\eqref{eq:drive}. In the next subsection, we discuss the intuition behind the introduction of the static parameters $\Omega_1$, $\Omega_4$.

\subsection{Constant part}

The Hamiltonian of Eq.~\eqref{eq:fguess} has a form similar to the Hamiltonian of Eq.~\eqref{eq:Hdriven}, except that it also involves two-qubit operators not conserving the total spin magnetization, $\sum_j \hat{Z}_j$. If one is experimentally able to modulate magnetization-nonconserving couplings like $\hat{X}\hat{X}$ and $\hat{Y}\hat{Y}$ separately, then the ansatz~\eqref{eq:fguess} would be sufficient. However, the Hamiltonian~\eqref{eq:Hdriven}, as compared to Eq.~\eqref{eq:fguess}, also contains terms $\hat{Y}_1\hat{Y}_2$ and $\hat{Y}_3 \hat{Y}_4$, which would lead to the formation of unwanted effective terms $\hat{Y}_1\hat{Z}_2\hat{Z}_3\hat{Y}_4$, in addition to $\hat{X}_1\hat{Z}_2\hat{Z}_3\hat{X}_4$. These terms are undesired since they do not commute with $\hat{X}_1\hat{Z}_2\hat{Z}_3\hat{X}_4$ on different plaquettes, and they are thus detrimental. Their suppression in the effective Hamiltonian can be achieved by introducing single-spin $\hat{X}$ drives on spins 1 and 4. The Hamiltonian \eqref{eq:fguess}, including the additional $\hat{Y}\hat{Y}$ terms, then becomes
\begin{align} 
& \Omega_1 \hat{X}_1 + \Omega_4 \hat{X}_4 + g_{13}(t) (\hat{X}_1 \hat{X}_3+\hat{Y}_1\hat{Y}_3)\nonumber \\
& + g_{23}(t) (\hat{X}_2\hat{X}_3 + \hat{Y}_2\hat{Y}_3) + g_{24}( t) (\hat{X}_2 \hat{X}_4+\hat{Y}_2\hat{Y}_4), \label{eq:Hguess2}
\end{align}
By choosing $\Omega_1=\Omega_4=M\omega$ with integer $M$ and moving to a rotating frame with respect to the $\hat{X}_1$ and $\hat{X}_4$ terms, the Hamiltonian transforms to
\begin{align}
&  g_{13}(t) (\hat{X}_1\hat{X}_2 + e^{i M\omega t\hat{X}_1} \hat{Y}_1 e^{-i M\omega t\hat{X}_1}\hat{Y}_2) \nonumber \\
& + g_{23}( t) (\hat{X}_2 \hat{X}_3 + \hat{Y}_2 \hat{Y}_3) \nonumber \\
& + g_{24}( t) (\hat{X}_2 \hat{X}_4 + \hat{Y}_2 e^{i M\omega t\hat{X}_4} \hat{Y}_4 e^{-i M\omega t\hat{X}_4})\ .
\end{align}
Note that, thanks to the choice of $\Omega_1$ and $\Omega_4$ as a multiple of $\omega$, the rotating frame coincides with the non-rotating frame at $t=\tau$. 
One can now observe that the terms multiplied by exponential functions oscillate at frequencies $\sim \pm (M\pm k)\omega$ with $k=1,2$, and will thus produce terms in a high-frequency expansion that are suppressed at least by a factor $1/(M\pm k)\omega$. By choosing $M$ sufficiently large, these terms can be made smaller than the effective four-body $\hat{X}_1\hat{Z}_2\hat{Z}_3\hat{X}_4$ term. A very large $M$ can become impractical, and this is an additional reason for including the parameters $\Omega_j$ in the optimization. 

Finally, the Hamiltonian of Eq.~\eqref{eq:Hguess2} will likely generate, at higher orders in the effective Hamiltonian, terms composed of the same operators which enter the undriven Hamiltonian itself. In the numerical optimization, small static corrections $\bar{g}_{ij}$ to the drive functions $g_{ij}(t)$ can also be included as additional free parameters, such that $g_{ij}(t)=\bar{g}_{ij} + g_{ij}\cos(n\omega t)$, for the purpose of cancelling these terms and permitting an additional fine tuning of the four-spin effective Hamiltonian. However, since the latter is already very clean via optimization of the oscillating components $g_{ij}$ (as shown in Sec.~\ref{sec:single_plaquette}), these corrections are of order $\sim10^{-4}\omega$ and thus beyond practical experimental interest.

\subsection{Effective reduction of the dynamical Lie algebra via time-symmetric drive} \label{app:lie}

In this subsection, we describe how the choice of time-symmetric drives, together with the specific algebraic structure of the control problem, leads to a situation where not all operators belonging to the dynamical Lie algebra can actually contribute to the effective Hamiltonian. The latter property is particularly advantageous, since it drastically reduces the number of potential unwanted operators, while easing the numerical optimization of parameters.

As the first step, we observe that the dynamical Lie algebra $\mathfrak{L}$, spanned by the operators in Table~\ref{tab:table1}, admits a Cartan decomposition~\cite{Dalessandro2007} $\mathfrak{L} = \mathfrak{h}\oplus\mathfrak{p}$, such that
\begin{equation} \label{eq:cartan}
[\mathfrak{h}, \mathfrak{h}] \subseteq \mathfrak{h}, \quad [\mathfrak{h}, \mathfrak{p}]\subseteq \mathfrak{p}, \quad [\mathfrak{p}, \mathfrak{p}] \subseteq \mathfrak{h} \ ,
\end{equation}
with subspaces $\mathfrak{h}$ and $\mathfrak{p}$ spanned by
\begin{align}
\mathfrak{h} = \mathrm{span}\{& \hat{Z}_1 \hat{Y}_2, \, \hat{Y}_3 \hat{Z}_4,\, \hat{X}_1\hat{Z}_2 \hat{Y}_3,\,  \hat{Y}_1 \hat{Z}_2 \hat{X}_3, \, \nonumber \\
& \hat{X}_2\hat{Z}_3\hat{Y}_4, \, \hat{Y}_2 \hat{Z}_3 \hat{X}_4,\,  \hat{Z}_1\hat{Z}_2\hat{Z}_3\hat{Y}_4, \nonumber \\
& \hat{Y}_1 \hat{Z}_2 \hat{Z}_3 \hat{Z}_4\} \ ,  \\
\mathfrak{p} = \mathrm{span} \{ & \hat{X}_1, \, \hat{X}_4,\, \hat{X}_1 \hat{X}_2, \nonumber \\
& \hat{X}_2 \hat{X}_3, \, \hat{X}_3 \hat{X}_4, \   \hat{Z}_1 \hat{Z}_2 \hat{Z}_3 \hat{Z}_4,\nonumber\\
 & \hat{Y}_1\hat{Y}_2, \, \hat{Y}_3\hat{Y}_4, \, \hat{Z}_1 \hat{Z}_2\hat{X}_3,\, \hat{X}_2 \hat{Z}_3 \hat{Z}_4, \nonumber \\
 & \hat{X}_1 \hat{Z}_2 \hat{Z}_3 \hat{X}_4,\, \hat{Y}_1 \hat{Z}_2 \hat{Z}_3 \hat{Y}_4 \} \ .
\end{align}

The dynamics $\hat{U}(t)$ of the system governed by the Hamiltonian $\hat{H}(t)$ can be formally expressed via a Magnus expansion~\cite{Blanes2009}, $\hat{U}(t) = e^{\hat{M}(t)}$, characterized by the Magnus exponent $\hat{M}(t) = \sum_{n=1}^\infty \hat{M}_n(t)$. The first terms read
\begin{align}
\hat{M}_1(t)&  = -i\int_0^t \! dt' \ \hat{H}(t')  \ , \\
\hat{M}_2(t) & = \frac{(-i)^2}{2} \ \int_0^{t} \! dt_1 \int_0^{t_1}\!\! dt_2 \ [\hat{H}(t_1), \hat{H}(t_2)]\ .
\end{align}
In particular, the $n$th term $\hat{M}_n(t)$ in the Magnus exponent involves $n-1$ nested commutators of $\hat{H}(t)$,
\begin{equation}
\Big[ \hat{H}(t_1),\big[\hat{H}(t_2), \dots[\hat{H}(t_{n-1}), \hat{H}(t_n)]\dots\big]\Big] \ ,
\end{equation}
evaluated at different times. Since $\hat{H}(t)$ is a linear combination of elements of $\mathfrak{p}$, it holds that $\hat{H}(t) \in \mathfrak{p}$. Hence, from the properties~\eqref{eq:cartan} of the Cartan decomposition, one can conclude by induction that
\begin{equation}
\hat{M}_{2n}(t) \in \mathfrak{h}, \quad \hat{M}_{2n+1}(t) \in \mathfrak{p}\ .
\end{equation}
Finally, choosing driving functions that are time-symmetric within the time interval $\tau$ makes even terms $\hat{M}_{2n}(\tau)$ in the Magnus expansion to vanish~\cite{Burum1981, Brinkmann2016}, as mentioned in Sec.~\ref{sec:oscpart}. As a result, the effective Hamiltonian is restricted to be a linear combination of operators belonging to $\mathfrak{p}$ only, rather than to the whole Lie algebra $\mathfrak{L}$. This can also be interpreted as the fact that a time-symmetric drive within the period cannot produce time-reversal-symmetry-breaking terms in the effective Hamiltonian~\cite{Hauke2012}.

\subsection{Numerical optimization}
\label{sec:numerical_optimization}

The numerical optimization is performed by starting with different random choices of drive parameters, computing the end-of-period propagator $\hat{U}_{ij}(\tau)$ generated by the driven dynamics and minimizing the norm
\begin{equation}
\Vert \hat{U}_{ij}(\tau) - e^{i \tau \mathcal{J} \hat{P}_{i,j}}\Vert 
\end{equation} 
with a gradient-descent algorithm.
Well converged results are obtained by first setting $\Omega_\alpha=0$ and optimizing $g_{\alpha\beta}$. Then, all parameters are optimized together starting from initial random guesses that are stochastically perturbed variants of the optimal $g_{\alpha\beta}$ found at the first optimization step. 

\section{Master equations} \label{app:master_equations}

In this Appendix, the master equations used to describe decoherence and dissipation, for both the undriven and driven four-plaquette system of Section~\ref{sec:single_plaquette}, are derived and discussed. While for the case of the undriven system the standard quantum-optical master equation~\cite{Breuer2007} applies, describing the thermal relaxation of each spin, the case of the driven system within the Floquet-Born-Markov formalism~\cite{Blumel1991, Grifoni1998, Hone2009, Mori2022} must be worked out explicitly due to the highly degenerate structure of the Floquet quasienergy spectrum for the four-spin plaquette, as discussed in the following.

The Hamiltonian of the system has the form of Eq.~\eqref{eq:Hschro} and can thus be written as $\hat{H}(t)=\hat{H}_0 + \hat{H}_d(t)$, where $\hat{H}_0=\sum_\alpha \omega_\alpha \hat{Z}_\alpha/2$ describes the bare spin energies, $\hat{H}_d(t)$ is the driven part, and the index $\alpha$ indicates coordinates in the lattice. Incoherent effects are modelled in the following through the coupling of each spin to an independent thermal bosonic bath at low temperature. 
The combined system-bath Hamiltonian $\hat{H}_{SB}(t)$ reads
\begin{equation}
\hat{H}_{SB}(t) = \hat{H}_0 + \hat{H}_d(t) + \hat{H}_B + \hat{H}_I\ ,
\end{equation}
where $\hat{H}_B$ is the bare Hamiltonian of the baths and where the system-bath interaction Hamiltonian $\hat{H}_I$ reads
\begin{equation}
 \hat{H}_I = \lambda \sum_{\alpha} \hat{X}_\alpha \otimes \hat{B}_\alpha,
 \end{equation}
 assuming equal system-bath coupling $\lambda$ for all spins, and where $\hat{B}_\alpha=\hat{B}_\alpha^\dagger$ are coupling operators for the $\alpha$th bath. Indicating with $\mathrm{tr}_B$ the partial trace over the baths' degrees of freedom and with $\hat{\rho}_B$ the baths density matrix, the thermal bosonic baths are characterized by the spectrum of the correlation functions~\cite{Breuer2007}
 \begin{align}
\gamma_\alpha(\omega) = &\int_{-\infty}^{+\infty} \!\! ds \ \mathrm{tr}_B\big[ e^{i\hat{H}_B s}\hat{B}_\alpha e^{-i\hat{H}_B s} \hat{B}_\alpha \hat{\rho}_B \big] e^{-i\omega s}, \nonumber \\
= &  \begin{cases}
J(\omega) n_\beta(\omega)
 & \omega \ge 0,\\
 J(-\omega) [1+n_\beta(-\omega)]
& \omega < 0,
\end{cases}
\label{eq:corrfun}
  \end{align}
where $n_\beta(\omega) = [e^{\beta \omega}-1]^{-1}$ denotes the Bose-Einstein distribution and where we choose an Ohmic spectral density $J(\omega) = \omega$. \\

\noindent {\bf Undriven system.} In the case of the undriven system, $\hat{H}_d(t)=0$, following standard derivations~\cite{Breuer2007}, one obtains the quantum-optical master equation for the system's density matrix $\hat{\rho}(t)$,
\begin{align}
\frac{d\hat{\rho}(t)}{dt} = -i[\hat{H}_0, \hat{\rho}(t)] &+ \lambda^2 \sum_\alpha \gamma(-\omega_\alpha) \mathcal{D}[\hat{\sigma}_\alpha^-]\hat{\rho}(t) \\
& +\lambda^2 \sum_\alpha  \gamma(\omega_\alpha) \mathcal{D}[\hat{\sigma}_\alpha^+]\hat{\rho}(t) \ ,
\end{align}
where $\mathcal{D}[\hat{c}]\hat{\rho}(t) = \hat{c}\hat{\rho}(t)\hat{c}^\dagger -\frac{1}{2}\hat{c}^\dagger \hat{c}\hat{\rho}(t)- \frac{1}{2}\hat{\rho}(t) \hat{c}^\dagger \hat{c}$ is a dissipator in Lindblad form.\\

\noindent{\bf Driven system.} As discussed in Section~\ref{sec:drive_ham} and Appendix~\ref{sec:intpicH}, the periodic Floquet protocol takes place in the interaction picture with respect to the undriven part $\hat{H}_0$, such that the time-periodic Hamiltonian is $\hat{H}(t) = \exp(i \hat{H}_0 t) \hat{H}_d(t)\exp(-i \hat{H}_0 t) = \hat{H}(t+\tau)$ with period $\tau=2\pi/\omega$. According to Floquet theorem~\cite{Shirley1965, Sambe1973, Eckardt2015}, the propagator generated by $\hat{H}(t)$ can be written in the form 
\begin{equation}
\hat{U}(t) = \sum_j e^{-i q_j t}\ket{u_j(t)}\!\bra{u_j(0)} = \hat{K}(t)e^{-it \hat{H}_F },
\end{equation}
where $q_j$ are the quasienergies, $\ket{u_j(t)}=\ket{u_j(t+\tau)}$ are the periodic Floquet modes, and where we have introduced the Floquet Hamiltonian $\hat{H}_F$, generating the stroboscopic dynamics, and the periodic micromotion operator $\hat{K}(t)=\hat{K}(t+\tau)$, which are defined by
\begin{subequations}
\begin{align}
&\hat{H}_F =\sum_j q_j\ket{u_j(0)}\!\bra{u_j(0)},\\
 & \hat{K}(t) = \sum_j \ket{u_j(t)}\!\bra{u_j(0)} .
\end{align}
\end{subequations}
For the single-plaquette Floquet engineering protocol of Section~\ref{sec:single_plaquette}, whose open dynamics is studied, the Floquet Hamiltonian (depicted in Fig.~\ref{fig:Figure2}) approximates very closely the plaquette operator $\hat{P}=-\mathcal{J} \hat{X}_1\hat{Z}_2\hat{Z}_3\hat{X}_4$ (using the labelling of Section~\ref{sec:single_plaquette} and Fig.~\ref{fig:Figure2}). The spectrum of the latter features two eightfold degenerate eigensubspaces, corresponding to eigenvalues $\langle \hat{P}\rangle = \pm \mathcal{J}$. The Floquet Hamiltonian is thus well approximated by
$\hat{H}_F = \mathcal{J}[\hat{\Pi}_+ - \hat{\Pi}_-]$ where $\hat{\Pi}_\pm$ represent the projectors into the $\pm\mathcal{J}$ degenerate subspaces, respectively. The propagator can, thus, be expressed as 
\begin{equation}
\hat{U}(t) = \hat{K}(t) [e^{-i\mathcal{J} t}\hat{\Pi}_+ + e^{i\mathcal{J} t}\hat{\Pi}_- ].
\end{equation}

In the basis of time-dependent Floquet modes for the system and in interaction picture for the bath, the system-bath interaction Hamiltonian reads
\begin{equation}
\hat{H}_{I}(t) = \lambda \sum_\alpha \hat{X}_\alpha(t)\otimes \hat{B}_\alpha(t)\ ,
\end{equation}
 where $\hat{B}_\alpha(t) = \exp(i\hat{H}_B t ) \hat{B}_\alpha\exp(-i\hat{H}_B t ) $ and where the system part is 
\begin{align}
\hat{X}_\alpha(t)& =  \hat{U}^\dagger(t) e^{i\hat{H}_0 t} \hat{X}_\alpha e^{-i\hat{H}_0 t}\hat{U}(t) \nonumber \\
& = \sum_{n,n'=\pm} \Big( e^{i (q_n - q_{n'} + \omega_\alpha) t}  \hat{\sigma}_{\alpha, nn'}^+(t)+ \mathrm{H.c.}\Big) \label{eq:Xat}
\end{align}
with $\hat{\sigma}_{\alpha,nn'}^+(t) = \hat{\Pi}_n\hat{K}^\dagger(t) \hat{\sigma}_{\alpha}^+\hat{K}(t)\hat{\Pi}_{n'}$ and $q_{\pm} = \pm \mathcal{J}$. Expanding $\hat{\sigma}_{\alpha}^+(t)$ in a Fourier series,
\begin{equation}
\hat{\sigma}_{\alpha}^+(t) =\sum_{m=-\infty}^{+\infty} \hat{\sigma}_{\alpha}^{+,(m)} e^{im\omega t},
\end{equation}
where the Fourier components satisfy $[\hat{\sigma}_{\alpha, nn'}^{+,(m)}]^\dagger = \hat{\sigma}_{\alpha, n'n}^{-,(-m)}$, one obtains that Eq.~\eqref{eq:Xat} features time-independent operators only, with the time dependence made fully explicit in the associated phase factors,
\begin{equation}  \label{eq:Xa_1}
\hat{X}_\alpha(t) =  \sum_{n,n'=\pm} \sum_{m=-\infty}^{+\infty} \Big( e^{i \Delta_{\alpha,nn'}^{(m)} t} \hat{\sigma}_{\alpha, nn'}^{+,(m)}+\mathrm{H.c.}\Big),
\end{equation}
which oscillate at frequencies $\Delta_{\alpha, nn'}^{(m)} = \omega_\alpha + (q_n - q_{n'}) + m\omega$.
Under Born-Markov approximations, the open dynamics of the system in described by the Markovian master equation~\cite{Breuer2007}
\begin{equation} \label{eq:markovme}
\frac{d\hat{\rho}(t)}{dt} = -\int_0^{+\infty} ds \mathrm{tr}_B[\hat{H}_I(t), [\hat{H}_I(t-s), \hat{\rho}(t)\otimes \hat{\rho}_B],
\end{equation}
where $\hat{\rho}(t)$ and $\hat{\rho}_B$ are the system and bath density matrices, respectively. 
Inserting Eq.~\eqref{eq:Xa_1} into Eq.~\eqref{eq:markovme}, one arrives at 
\begin{widetext}
\begin{align}
\frac{d\hat{\rho}(t)}{dt} = \lambda^2 \sum_\alpha\sum_{\substack{i,i',\\ j,j'=\pm}}\sum_{m,m'=-\infty}^{+\infty} e^{i (\Delta_{\alpha, ii'}^{(m)}-\Delta_{\alpha, jj'}^{(m')})t}\Big\{\Big[ \big( \hat{\sigma}_{\alpha, jj'}^{+, (m')}\big)^\dagger \hat{\rho}(t)\hat{\sigma}_{\alpha, ii'}^{+,(m)}- \hat{\sigma}_{\alpha, ii'}^{+,(m)}\big( \hat{\sigma}_{\alpha, jj'}^{+, (m')}\big)^\dagger \hat{\rho}(t)\Big] G_\alpha(-\Delta_{\alpha, jj'}^{(m)})\\
+ \Big[ \hat{\sigma}_{\alpha, ii'}^{+, (m)} \hat{\rho}(t)\big(\hat{\sigma}_{\alpha, jj'}^{+,(m')}\big)^\dagger -  \hat{\rho}(t)\big(\hat{\sigma}_{\alpha, jj'}^{+,(m')}\big)^\dagger \hat{\sigma}_{\alpha, ii'}^{+, (m)} \Big] G^*_\alpha(\Delta_{\alpha, jj'}^{(m)})\Big\} + \mathrm{H.c.}, 
\end{align}
\end{widetext}
where $G_\alpha(\omega)$ is the one-sided Fourier transform of the bath correlation functions
\begin{equation}
G_{\alpha}(\omega) = \int_0^{+\infty} \!\! \! ds \ e^{-i\omega s} \mathrm{tr}_B [\hat{B}_\alpha(s) \hat{B}_\alpha(0) \hat{\rho}_B].
\end{equation}
Under the assumption of weak system-bath coupling, the secular (rotating-wave) approximation is adopted: only terms such that $\Delta_{\alpha, ii'}^{(m)}-\Delta_{\alpha, jj'}^{(m')}=(q_i - q_{i'})- (q_j - q_{j'}) + (m-m')\omega = 0$, while others are neglected. While the hierarchy $|q_i-q_{i'}| \le 2\mathcal{J} \ll \omega$ justifies the so-called `moderate' rotating-wave approximation, $m=m'$, the full secular approximation (implying that $\Delta_{\alpha, ii'}^{(m)}-\Delta_{\alpha, jj'}^{(m')}$ vanishes only if $ i=j$, $i'=j'$ and $m=m'$) cannot be applied. Indeed, given the quasienergy structure of the Floquet Hamiltonian, composed of two degenerate `bands' of states, the following combinations of indices must be considered instead
\begin{align}
\{ii', jj'\} = \big\{& (++,++),\ (++, --),\ (+-, +-), \nonumber \\
& (-+,-+),\
(--, ++),\ (--, --) \big\}.
\end{align}
Using the explicit expressions
\begin{subequations}
 \begin{align}
& \Delta_{\alpha,++}^{(m)}=\Delta_{\alpha,--}^{(m)}= \omega_\alpha + m\omega,\\
& \Delta_{\alpha,+-}^{(m)}= \omega_\alpha + 2\mathcal{J} +  m\omega,\\
& \Delta_{\alpha,-+}^{(m)}= \omega_\alpha - 2\mathcal{J} +  m\omega,
\end{align}
\end{subequations}
using Eq.~\eqref{eq:corrfun} and neglecting Lamb shift contributions, one finally obtains the master equation in Lindblad form
\begin{widetext}
\begin{align} 
\frac{d\hat{\rho}(t)}{dt} =\lambda^2 \sum_\alpha  \sum_{m=-\infty}^{+\infty} \Big\{ & \gamma_\alpha(-\omega_\alpha+m\omega) \mathcal{D}\big[\hat{\sigma}_{\alpha, ++}^{-,(m)}+\hat{\sigma}_{\alpha, --}^{-,(m)}\big]\hat{\rho}(t) + \gamma_\alpha\big(-\omega_\alpha - 2\mathcal{J} +m\omega\big)\mathcal{D}\big[\hat{\sigma}_{\alpha, -+}^{-,(m)}\big]\hat{\rho}(t)\nonumber \\
&+ \gamma_\alpha\big(-\omega_\alpha + 2\mathcal{J} + m\omega\big) \mathcal{D}\big[\hat{\sigma}_{\alpha, +-}^{-,(m)}\big]\hat{\rho}(t) +\gamma_\alpha(\omega_\alpha+m\omega) \mathcal{D}\big[\hat{\sigma}_{\alpha, ++}^{+,(m)}+\hat{\sigma}_{\alpha, --}^{+,(m)}\big]\hat{\rho}(t)\nonumber  \\
& + \gamma_\alpha\big(\omega_\alpha + 2\mathcal{J} + m\omega \big)\mathcal{D}\big[\hat{\sigma}_{\alpha, +-}^{+,(m)} \big]\hat{\rho}(t)
+ \gamma_\alpha\big(\omega_\alpha - 2\mathcal{J} + m\omega\big)\mathcal{D}\big[\hat{\sigma}_{\alpha, -+}^{+,(m)} \big]\hat{\rho}(t) \Big\}\ .
\label{eq:me_floquet}
\end{align}
\end{widetext}
This master equation has a simple physical interpretation in terms of transitions within or between Floquet `bands' occurring through the excitation ($\hat{\sigma}_\alpha^+$) or relaxation ($\hat{\sigma}_\alpha^-$) of the $\alpha$th spin, which can be assisted by the Floquet drive via the exchange of an integer number of drive quanta $m\omega$. These processes have energy cost $\pm\omega_\alpha + m\omega$ for the spin flip with potential intra-Floquet-band transitions, and they involve an additional energy $\pm 2\mathcal{J}$ for changing Floquet band. For example, the second term in the first line of Eq.~\eqref{eq:me_floquet} describes transitions from the upper ($+$) to the lower ($-$) Floquet band via the coupling operator $\hat{\sigma}_\alpha^-$, assisted by the emission of $m$ quanta $\omega$ in the drive. In the process, the system thus releases the excess energy $\omega_\alpha + 2\mathcal{J} -m\omega$ to the environment.

\section{Semi-analytical representation of operators} 
\label{sec:cpr_representation}

In order to efficiently represent and manipulate the operators for large systems in numerical simulations, in particular for computing commutators, we employ the following semi-analytical representation. 
A generic multi-qubit operator
\begin{equation}
\hat{O} =\hspace{-0.5cm} \sum_{\substack{\hat{\sigma}_{\alpha_j} \in \\ \{\hat{X}_{\alpha_j}, \hat{Y}_{\alpha_j}, \hat{Z}_{\alpha_j},\hat{\mathbb{1}}_{\alpha_j}\}}} \hspace{-0.5cm} O_{\sigma_{\alpha_1},\dots,\sigma_{\alpha_N}} \hat{\sigma}_{\alpha_1} \otimes\dots \otimes \hat{\sigma}_{\alpha_N},
\end{equation}
is represented in a sparse-like form by storing (i) the position of non-identity terms, (ii) the type of Pauli matrix at each position (as a string), (iii) the coefficient $O_{\sigma_{\alpha_1},\dots,\sigma_{\alpha_N}} $. For example, an operator $\hat{O}=\omega_1 \hat{Z}_1 + \omega_2 \hat{Z}_2 + J \hat{X}_1 \hat{X}_2$ is represented as
\begin{align}
O = \{&[1\to `Z\text{'},  \omega_1], \nonumber \\
&[2\to `Z\text{'},  \omega_2], \nonumber \\
&[(1, 2) \to (`X\text{'}, `X\text{'}),  J]\}\ .
\end{align}
This representation is convenient for manipulating Hamiltonians for large systems since it allows one to perform matrix multiplications without needing to store and multiply the matrices explicitly. This is particularly advantageous in the case of Floquet-engineered effective Hamiltonians, since they are typically not sparse, because of all the higher-order effective terms produced by the Floquet drives. For example, the commutator of two operators $\hat{O}_1$ and $\hat{O}_2$ is computed by simply selecting shared non-identity sites between $O_1$ and $O_2$ and providing the analytical result of the commutator of such terms. 

\section{Construction of the ground-state preparation protocol} \label{app:gs_prot}

The aim of this Section is to explain how the ground state preparation protocol proposed in Sec.~\ref{sec:gs_prep} can indeed produce the same results of Eq.~\eqref{eq:gs_proj}. This is based on three observations:

1) Since $\hat{X}$ and $-i\hat{Y}$ give the same result if applied to $\ket{0}$, $\hat{X}\ket{0}=-i \hat{Y} \ket{0}=\ket{1}$, it is also true that 
\begin{equation}
(1+\hat{P}_{i,j})\ket{\psi_0} = (1-i \hat{A}_{i,j})\ket{\psi_0}=(1-i \hat{B}_{i,j})\ket{\psi_0} \ .
\end{equation} 

2) $\hat{A}_{i,j}$ commutes with all other operators $\hat{A}_{i',j'}$ and $\hat{B}_{i',j'}$ except for $\hat{A}_{i+1,j+1}$ and $\hat{B}_{i-1,j-1}$; this implies that the ordering of the product along diagonals spanning the system in direction bottom-right (BR) to top-left (TL) does not matter, whereas the ordering of products along diagonals in direction bottom-left (BL) to top-right (TR) does matter. The correct order can be determined from the next point.

3) In a product of two noncommuting terms $(1-i\hat{A}_{i+1,j+1})$ and $(1-i\hat{A}_{i,j})$, the noncommuting parts satisfy
\begin{align}
-\hat{A}_{i+1,j+1} \hat{A}_{i,j} =& - i  \hat{X}_{i,j} \hat{Z}_{i+1,j} Z_{i, j+1} \hat{Z}_{i+1, j+1} \hat{Z}_{i+1,j+1} \nonumber\\
& \otimes \hat{Z}_{i+2,j+1} \hat{Z}_{i+1, j+2} \hat{Y}_{i+2, j+2}\ .\label{eq:noncomm1}
\end{align}
According to observation 1 and the fact that $\hat{Z}\ket{0}=\ket{0}$, the operator in Eq.~\eqref{eq:noncomm1} acts on $\ket{\psi_0}$ in exactly the same way as $\hat{P}_{i+1,j+1} \hat{P}_{i,j}$, such that
\begin{multline}
(1-i \hat{A}_{i+1,j+1})(1-i\hat{A}_{i,j})\ket{\psi_0} =\\ (1+\hat{P}_{i+1,j+1})(1+\hat{P}_{i,j})\ket{\psi_0}.
\end{multline}
Generalizing, the product of $n$ terms $-i \hat{A}_{i,j}$ along a diagonal ordered from BL to TR has the same action on $\ket{\psi_0}$ as the equivalent product of $\hat{P}_{i,j}$.

\end{document}